# Time-Crystal Particles and Classical Spin 4-vector

Mário G. Silveirinha[*]

[(1)] *University of Lisbon–Instituto Superior Técnico and Instituto de Telecomunicações, Avenida Rovisco Pais, 1, 1049-001 Lisboa, Portugal,* [mario.silveirinha@co.it.pt](mario.silveirinha@co.it.pt)

**Abstract**

Time crystals are exotic phases of matter characterized by a broken time-translational symmetry, such that the ground state of the system evolves in time in a periodic fashion. Even though the time-crystal concept was introduced relatively recently, a related proposal can be traced back to Louis de Broglie. In his thesis, de Broglie conjectured that elementary particles may have some sort of internal clock that rules their behavior in the microscopic world. Here, I revisit de Broglie's idea and demonstrate that a special extreme case of classical mechanics yields in a natural way time-crystal type dynamics. Remarkably, it is found that time-crystal particles are characterized by a spin 4-vector that has a purely kinematic origin and is determined by the binormal of the velocity trajectory in the Bloch sphere. The dynamics of time-crystal particles is ruled by a generalized least action principle, such that the particle dynamically probes the nearby space and moves on average towards the direction that minimizes the action. I apply the theory to the case of charged particles and find that it predicts spin precession and that it provides a simple and intuitive picture for the origin of the anomalous magnetic moment of the electron. The classical formalism is recovered as an "effective theory" valid on a coarse time scale.

---

[*] To whom correspondence should be addressed: E-mail: [mario.silveirinha@co.it.pt](mario.silveirinha@co.it.pt)

# I. Introduction

During the last decade there has been a great interest in the physics of "time-crystals" [1-6]. A time-crystal has a spontaneously broken time-translational symmetry such that its ground-state evolves periodically in time, notwithstanding the equations of motion are invariant under arbitrary time-translations (i.e., do not depend on the time origin). Thus, the ground state of a time-crystal may be regarded as some sort of "perpetuum mobile" that oscillates indefinitely with some time period [1]. The time-crystal terminology was introduced by Wilczek [4].

Remarkably, "time-crystal" concept can be traced back to the very beginning of quantum mechanics and to Louis de Broglie, who almost one hundred years ago in his dissertation conjectured that an electron has an internal clock that ticks with an angular frequency $mc^2/\hbar$ [7-9]. He suggested that the angular momentum quantization has its origin in some unknown internal periodic process. His ideas set the stage for the development of wave mechanics.

Following de Broglie's steps, it is natural to wonder if time-crystals can play a role in the explanation of physical phenomena at a fundamental level. Motivated by the link between de Broglie's clock and time-crystals, here I investigate a "time-crystal" model of particles that arises naturally from an extreme case of classical mechanics: a particle with vanishing mass $m \to 0$.

As is well-known, the classical equations become ill-defined in the $m \to 0$ limit, because the particle speed becomes identical to the speed of light $c$ and the Lorentz factor diverges to infinity, which are properties seemingly deprived of physical sense. Here, I show that there is a simple path to fix the singularities of the $m \to 0$ limit, which leads naturally to a time-crystal type (Lorentz-covariant) dynamics. Interestingly, the proposed model predicts that "time-crystal" particles are characterized by a spin 4-vector associated with the incessant spinning motion of the particle. The spin 4-vector has a purely kinematic origin and is determined by the binormal of the velocity trajectory in the Bloch sphere.

Furthermore, it is demonstrated that the trajectory of a time-crystal particle is controlled by a *dynamical* least action principle, which generalizes the classical action. Due of the incessant spinning motion, the time-crystal particle dynamically probes every direction of space and moves on average towards the direction of space that minimizes a generalized action integral. For closed orbits, the generalized action is Lorentz invariant and quantized in units of $2\pi\hbar$.

The developed model predicts that mass is an emergent property and that a time-crystal particle is formed by two components: one component behaves as some sort of "pilot" wave that probes the nearby space at the speed of light, whereas the other component behaves as a standard relativistic massive particle. Thus, the time-crystal formalism predicts a peculiar form of wave-particle duality and is reminiscent of the pilot-wave description of quantum mechanics introduced by de Broglie and Bohm [10-11].

This article is organized as follows. Section II introduces the basic ideas of the proposed time-crystal model. Starting from the study of the relativistic kinematics of a massless accelerated point particle, I introduce a spin 4-vector determined by the binormal of the velocity trajectory in the Bloch sphere. It is demonstrated that the time evolution of the time-crystal electron is controlled by an action integral and by a dynamical least action principle. In Sects. III and IV, I introduce the notion of center of mass frame and demonstrate that it naturally leads to the idea that a time-crystal particle is formed by two-components: the original (wave-type) component that moves at the speed of light, and a "center-of-mass" component that effectively behaves as a massive particle. Sections V and VI are devoted to the study of charged time-crystal particles (let us say a "time-crystal electron") with the electromagnetic field. The classical theory is recovered as an "effective theory" valid in the $\hbar \to 0$ limit. It is demonstrated that the time-crystal model predicts the spin vector precession in a magnetic field and provides an intuitive picture for the physical origin of the anomalous magnetic moment. Finally, Sect. VII provides a summary of the main results.

## II. Time-crystal particles and spin

In this section, it is demonstrated that massless particles have a time-crystal type dynamics that is controlled by a generalized least action principle. It is shown that a massless particle is characterized a classical spin 4-vector determined by the binormal of the velocity trajectory in the Bloch sphere.

### *A. Postulates*

The theory of classical massive particles relies on the Lagrangian formalism. The classical equations become ill-defined in the zero-mass limit ($m \to 0$) as Lorentz factor diverges to infinity ($\gamma \to \infty$). In fact, in the $m \to 0$ limit the energy-momentum dispersion becomes linear, which requires that the particle velocity has amplitude *c*. Thus, a free-particle with $m=0$ must travel at speed *c* in a straight line. Thereby, the limit $m \to 0$ is seemingly deprived of physical meaning, as the familiar of properties of classical objects are certainly incompatible with those of particles that travel at speed *c* in a straight line.

Let us however consider a theory for a "time-crystal" particle that is not directly based on the classical Lagrangian formalism but on the following two simple postulates (below $\boldsymbol{\beta}$ stands for the particle velocity normalized to *c* and $\dot{\boldsymbol{\beta}} = d\boldsymbol{\beta}/dt$ is a normalized acceleration):

- **P1** *The time-crystal particle speed is c for every inertial observer:* $|\boldsymbol{\beta}| = 1$.

- **P2** *The time-crystal particle acceleration never vanishes,* $\dot{\boldsymbol{\beta}} \neq 0$*, even for a "free" particle.*

I complement the above postulates with the requirement that the equations of motion must be Lorentz-covariant, i.e., the theory must be consistent with special relativity. The time-crystal particle will also be designated as the "wave" to highlight the fact that similar to a light-wave it travels with speed *c*.

The second postulate cannot be reconciled with the Hamiltonian formalism in any obvious way. Indeed, the condition $\dot{\boldsymbol{\beta}} \neq 0$ implies that the momentum ($\mathcal{P} = \frac{\mathcal{E}}{c}\boldsymbol{\beta}$) of a massless free particle must vary with time; thereby in a generic reference frame the (wave) energy $\mathcal{E}$ must also be time dependent. Thus, a theory compatible with the two postulates cannot rely on a (time-independent) Hamiltonian.

Let us analyze some immediate geometrical consequences of the postulates. First of all, since $\boldsymbol{\beta} \cdot \boldsymbol{\beta} = 1$ the acceleration and the velocity are forcibly perpendicular $\boldsymbol{\beta} \cdot \dot{\boldsymbol{\beta}} = 0$. Thus, the real-space trajectory of the time-crystal particle, $\mathbf{r} = \mathbf{r}_0(t)$, is necessarily curved (Fig. 1, left). Interestingly, the curvature of the real-space trajectory $K_R = \frac{|c\boldsymbol{\beta} \times c\dot{\boldsymbol{\beta}}|}{|c\boldsymbol{\beta}|^3}$ [12] is determined by the normalized acceleration:

$$K_R = \frac{|\dot{\boldsymbol{\beta}}|}{c}, \qquad \text{(curvature of the real-space trajectory)}. \tag{1}$$

A larger acceleration implies a more curved trajectory, with a smaller curvature radius $R = 1/K_R$. Note that since $\dot{\boldsymbol{\beta}} \neq 0$ the curvature cannot vanish, and hence the trajectory cannot be a straight line.

Besides the real-space trajectory, it is relevant to analyze the trajectory of the velocity vector. Since $\boldsymbol{\beta} \cdot \boldsymbol{\beta} = 1$ the normalized velocity may be regarded as a vector in the Bloch sphere (sphere with unit radius; see Fig. 1, right). As the velocity cannot be a constant vector, the trajectory $\boldsymbol{\beta}(t)$ in the Bloch sphere has necessarily some nonzero curvature:

$$K = \frac{|\dot{\boldsymbol{\beta}} \times \ddot{\boldsymbol{\beta}}|}{|\dot{\boldsymbol{\beta}}|^3} \neq 0, \quad \text{(curvature of the velocity trajectory)}. \tag{2}$$

Thus, both the real-space and the velocity trajectories are necessarily curved.

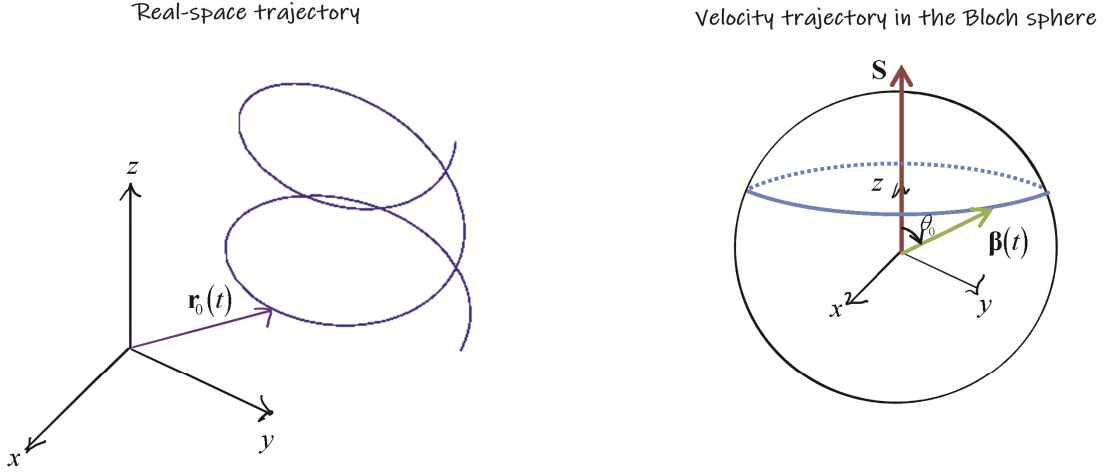

**Fig. 1 Left:** Sketch of the real-space trajectory $\mathbf{r}_0(t)$ of a time-crystal particle. **Right:** Sketch of the trajectory of the normalized velocity $\boldsymbol{\beta}(t)$ in the Bloch sphere. In the sketch, the trajectory in the Bloch sphere is planar (blue circle) and the spin vector $\mathbf{S}$ is perpendicular to the plane that contains the velocity trajectory. Both the real-space and the Bloch-sphere trajectories are necessarily curved for a particle satisfying the two postulates.

### B. Wave energy-momentum 4-vector

Since $\boldsymbol{\beta}$ transforms as a relativistic velocity under a Lorentz boost [13, Sects. A and B], one may construct with it an energy-momentum 4-vector $(\mathcal{E}, \mathcal{P})$ in a standard way. Specifically, the (wave) momentum is related to the velocity as

$$\mathcal{P} = \frac{\mathcal{E}}{c}\boldsymbol{\beta}, \tag{3}$$

with $\mathcal{E}$ the wave energy. The wave energy describes an interaction of the time-crystal particle with itself, and its physical meaning will be further elaborated in Sect. IV.C. Because of the first postulate, the wave energy-momentum dispersion is linear, $|\mathcal{E}| = c|\mathcal{P}|$, as expected for a massless particle.

Since $(\mathcal{E}, \mathcal{P})$ is a 4-vector it transforms under a Lorentz boost as:

$$\mathcal{E}' = \gamma(\mathcal{E} - \mathcal{P}\cdot\mathbf{v}_{\text{rel}}), \qquad \mathcal{P}' = \boldsymbol{\Gamma}\cdot\left(\mathcal{P} - \mathcal{E}\frac{\mathbf{v}_{\text{rel}}}{c^2}\right). \tag{4}$$

The primed quantities are measured in a (primed) inertial frame that moves with speed $\mathbf{v}_{rel}$ with respect to the unprimed reference frame. In the above, $\gamma = 1/\sqrt{1 - \mathbf{v}_{rel} \cdot \mathbf{v}_{rel}/c^2}$ is the Lorentz factor and $\mathbf{\Gamma}$ is the tensor $\mathbf{\Gamma} = (\gamma \hat{\mathbf{v}}_{rel} \otimes \hat{\mathbf{v}}_{rel} + (\mathbf{1} - \hat{\mathbf{v}}_{rel} \otimes \hat{\mathbf{v}}_{rel}))$ with $\hat{\mathbf{v}}_{rel} = \mathbf{v}_{rel}/|\mathbf{v}_{rel}|$ a unit vector. The symbol $\otimes$ represents the tensor product of two vectors. Using $\mathcal{P} = \frac{\mathcal{E}}{c}\boldsymbol{\beta}$ one sees that in the primed reference frame the wave energy can be written as $\mathcal{E}' = \gamma D \mathcal{E}$ with $D = 1 - \frac{\mathbf{v}_{rel}}{c} \cdot \boldsymbol{\beta}$. This formula shows that $\mathrm{sgn}(\mathcal{E}) = \mathrm{sgn}(\mathcal{E}')$, i.e., the sign of the energy is frame independent.

It is shown in the supplementary information [13, Sect. B] that the acceleration of a massless particle transforms under a Lorentz boost as:

$$\left|\dot{\boldsymbol{\beta}}'\right| = \frac{1}{\gamma^2}\frac{1}{D^2}\left|\dot{\boldsymbol{\beta}}\right|, \tag{5}$$

Hence, combining $\mathcal{E}' = \gamma D \mathcal{E}$ with the above formula one finds that for a massless particle

$$\mathcal{E}'^2 \left|\dot{\boldsymbol{\beta}}'\right| = \mathcal{E}^2 \left|\dot{\boldsymbol{\beta}}\right|, \tag{6}$$

i.e., $\left|\dot{\boldsymbol{\beta}}\right|\mathcal{E}^2$ is a Lorentz scalar and thereby has the same value in any inertial frame at corresponding spacetime points. The Minkowski product of any two 4-vectors is a Lorentz scalar. Thus, if $(E_1, \mathbf{p}_1)$ and $(E_2, \mathbf{p}_2)$ transform as 4-vectors, then $\mathcal{E}(E_i - c\mathbf{p}_i \cdot \boldsymbol{\beta})$ are Lorentz scalars. This property and Eq. (6) indicate that the normalized acceleration is a function of the form

$$\left|\dot{\boldsymbol{\beta}}\right| \sim (E_1 - c\mathbf{p}_1 \cdot \boldsymbol{\beta})(E_2 - c\mathbf{p}_2 \cdot \boldsymbol{\beta}), \tag{7}$$

for some $(E_i, \mathbf{p}_i)$. It will be shown in Sect. III.C that $(E_i, \mathbf{p}_i)$ are related to the classical canonical 4-momentum and to the classical massive 4-momentum.

### C. Spin 4-vector

Consider now a Frenet-Serret frame $\hat{\mathbf{t}}, \hat{\mathbf{n}}, \hat{\mathbf{b}}$ associated with the *velocity trajectory*, formed by tangent, normal and binormal vectors, respectively [12]. The Frenet-Serret frame is well defined because the velocity trajectory in the Bloch sphere cannot be a straight line ($K \neq 0$). From standard differential geometry, it is possible to write the Frenet-Serret basis explicitly as $\hat{\mathbf{t}} = \dot{\boldsymbol{\beta}}/|\dot{\boldsymbol{\beta}}|$, $\hat{\mathbf{b}} = \dot{\boldsymbol{\beta}} \times \ddot{\boldsymbol{\beta}}/|\dot{\boldsymbol{\beta}} \times \ddot{\boldsymbol{\beta}}|$ and $\hat{\mathbf{n}} = \hat{\mathbf{b}} \times \hat{\mathbf{t}}$. The dots represent time derivatives.

In the supplementary information [13, Sect. C], it is shown that the acceleration can be expressed as:

$$\dot{\boldsymbol{\beta}} = \Omega_s \mathbf{S} \times \boldsymbol{\beta}, \tag{8}$$

where $\Omega_s$ is a scalar function and $\mathbf{S}$ is a vector parallel to the binormal of the velocity trajectory. It is convenient to pick $\mathbf{S}$ equal to:

$$\mathbf{S} = \frac{1}{|\dot{\boldsymbol{\beta}}|^3} \left( \dot{\boldsymbol{\beta}} \times \ddot{\boldsymbol{\beta}} \right). \tag{9}$$

The vector $\mathbf{S}$ will be referred to as the *spin vector*. Quite remarkably (see [13, Sect. B]), for a massless particle, $(\mathbf{S} \cdot \boldsymbol{\beta}, \mathbf{S})$ transforms as a 4-vector under a Lorentz transformation, such that:

$$\mathbf{S}' = \boldsymbol{\Gamma} \cdot \left( \mathbf{S} - \frac{\mathbf{v}_{\text{rel}}}{c} \boldsymbol{\beta} \cdot \mathbf{S} \right), \qquad \boldsymbol{\beta}' \cdot \mathbf{S}' = \gamma \left( \boldsymbol{\beta} \cdot \mathbf{S} - \mathbf{S} \cdot \frac{\mathbf{v}_{\text{rel}}}{c} \right). \tag{10}$$

The tensor $\boldsymbol{\Gamma}$ is defined as in Sect. II.B. Thus, the binormal of the velocity trajectory (apart from a scaling factor) is the space component of a 4-vector. To our best knowledge, this result was not previously reported in the literature. The Lorentz scalar associated with the 4-vector is [13, Sect. B]:

$$\mathbf{S} \cdot \mathbf{S} - (\mathbf{S} \cdot \boldsymbol{\beta})^2 = |\mathbf{S} \times \boldsymbol{\beta}|^2 = 1. \tag{11}$$

The spin 4-vector has no dimensions and determines the axis of rotation of the velocity in the Bloch sphere (see Fig. 1, right). Evidently, $\dot{\boldsymbol{\beta}} \cdot \mathbf{S} = 0$ at any time instant.

Furthermore, combining Eqs. (8) and (9), one sees that $\Omega_s = |\dot{\boldsymbol{\beta}}|$, and hence $\Omega_s$ is necessarily a strictly positive function. The parameter $\Omega_s$ will be referred to as the "spin" angular frequency, as it links the speed of the particle ($c$) with the radius of the curvature ($R = 1/K_R$) of the real space trajectory: $c = \Omega_s R$ [Eq. (1)]. The spin frequency is exactly the (normalized) acceleration amplitude ($\Omega_s = |\dot{\boldsymbol{\beta}}|$).

Equation (8) determines the dynamical law for $\boldsymbol{\beta}$. Thereby, the variation in time of the velocity is controlled by the spin vector and by the spin frequency $\Omega_s$. It should be noted that Eq. (8) with $\Omega_s = |\dot{\boldsymbol{\beta}}|$ is manifestly Lorentz covariant, because the previous derivation holds true in any inertial frame due to the properties of the spin 4-vector.

From Eq. (2), the amplitude of the space component of the spin 4-vector gives precisely the curvature of the trajectory of the normalized velocity in the Bloch sphere:

$$K = |\mathbf{S}|. \tag{12}$$

For example, if the velocity trajectory lies in some circle in the Bloch sphere, the curvature $K$ is minimal ($|\mathbf{S}|=1$) for a great circle and is larger for smaller circles. In particular, the spin vector amplitude is minimal ($|\mathbf{S}|=1$) when $\mathbf{S}$ is perpendicular to the velocity.

On the other hand, the time component of the 4-vector, determines the angle $\theta$ between the spin vector and the normalized velocity in the Bloch sphere ($\frac{\mathbf{S}}{|\mathbf{S}|} \cdot \boldsymbol{\beta} = \cos\theta$; see Fig. 1, right, with $\theta = \theta_0 = const.$):

$$\mathbf{S} \cdot \boldsymbol{\beta} = K \cos\theta. \tag{13}$$

The time component of the 4-vector may also related to the torsion and curvature [Eq. (1)] of the real-space trajectory $\mathbf{r} = \mathbf{r}_0(t)$. The torsion of the real space trajectory is [12]:

$$\tau_R = \frac{\dot{\mathbf{r}}_0 \cdot (\ddot{\mathbf{r}}_0 \times \dddot{\mathbf{r}}_0)}{|\dot{\mathbf{r}}_0 \times \ddot{\mathbf{r}}_0|^2} = \frac{1}{c} \frac{\dot{\boldsymbol{\beta}} \cdot (\ddot{\boldsymbol{\beta}} \times \dddot{\boldsymbol{\beta}})}{|\dot{\boldsymbol{\beta}}|^2}. \tag{14}$$

Using Eqs. (1) and (9) it follows that the time component of the spin 4-vector is the ratio between the torsion and curvature of the real-space trajectory:

$$\mathbf{S} \cdot \boldsymbol{\beta} = \frac{\tau_R}{K_R}. \tag{15}$$

In particular, the real-space trajectory can be contained on a plane ($\tau_R = 0$) if and only if the time-component of the spin 4-vector vanishes.

## D. Integration of the equations of motion

In order to have some insight of the dynamics of a time-crystal particle, one may integrate formally $\dot{\boldsymbol{\beta}} = \Omega_s \mathbf{S} \times \boldsymbol{\beta}$ [Eq. (8)], regarding for now the spin angular frequency and $\mathbf{S}$ as known functions of time. This gives:

$$\boldsymbol{\beta}(t = t_N) = e^{dt\Omega_s(t_{N-1})\mathbf{S}(t_{N-1}) \times \mathbf{1}} .. e^{dt\Omega_s(t_1)\mathbf{S}(t_1) \times \mathbf{1}} e^{dt\Omega_s(t_0)\mathbf{S}(t_0) \times \mathbf{1}} \boldsymbol{\beta}(t_0), \tag{16}$$

where it is implicit that $t_0, t_1, ..., t_N$ are time instants separated by an infinitesimal amount $dt$. Clearly, the velocity dynamics is ruled by a sequence of infinitesimal rotations about the spin vector with instantaneous rotation frequency $\omega_s = \Omega_s |\mathbf{S}|$.

Let us focus on the particular case of a planar trajectory in real-space, for which the spin vector is necessarily a vector perpendicular to the plane of motion with $|\mathbf{S}| = 1$. In this case, it is possible to write:

$$\boldsymbol{\beta}(t) = e^{\varphi(t)\mathbf{S} \times \mathbf{1}} \boldsymbol{\beta}(t_0), \qquad \varphi(t) = \int_{t_0}^{t} \Omega_s(t') dt'. \tag{17}$$

As seen, $\boldsymbol{\beta}(t)$ differs from the initial velocity $\boldsymbol{\beta}(t_0)$ by a rotation of $\varphi(t)$ about the spin axis. For example, if the motion is confined to the *xoy* plane ($\hat{\mathbf{S}} = \hat{\mathbf{z}}$) the normalized velocity can be written explicitly as $\boldsymbol{\beta}(t) = (\cos(\varphi_0 + \varphi(t)), \sin(\varphi_0 + \varphi(t)), 0)$. Interestingly, for a planar motion $\boldsymbol{\beta}(t)$ can be identified with a complex number ($\boldsymbol{\beta}(t) \sim e^{i\varphi(t)} e^{i\varphi_0}$) and the rotation operator $e^{\varphi(t)\mathbf{S} \times \mathbf{1}}$ with a multiplication by $e^{i\varphi(t)}$.

Suppose now that $\Omega_s$ is approximately constant in the time interval of interest. Then, the real-space trajectory corresponds to a circumference with radius $c/\Omega_s$ (for simplicity, the origin is taken as the center of the trajectory; $t_0 = 0$ and $\varphi_0 = \pi/2$ ):

$$\mathbf{r}_0(t) = \frac{c}{\Omega_s}\left(\cos(\Omega_s t), \sin(\Omega_s t), 0\right). \tag{18}$$

Remarkably, notwithstanding the massless particle moves with speed *c*, there is *no net translational motion* in the considered reference frame: the time-crystal particle merely rotates around the spin vector with the angular velocity $\Omega_s$. Both the position vector $\mathbf{r}_0(t)$ and the velocity $\boldsymbol{\beta}(t)$ whirl around the same axis (parallel to the spin vector). The spinning motion determines some sort of "internal clock" or "de Broglie clock" that determines the periodicity in time of the particle state.

## E. *Generalized least action principle*

The most fundamental description of the dynamics of physical systems is based on an "action integral". The time evolution of the system state is such that it corresponds to a stationary point (typically a minimum) of the "action". Next, I show that the time evolution of a time-crystal particle is determined by a generalized action integral and by a dynamical least action principle.

For simplicity, I consider some generic planar orbit of the time-crystal particle. In this case, the trajectory of $\boldsymbol{\beta}$ in the Bloch-sphere is the great circle perpendicular to the spin vector. As seen in Sect. II.D, the trajectory of $\boldsymbol{\beta}$ in this great circle can be identified with a complex number $\boldsymbol{\beta} \sim e^{i\varphi}$. In particular, Eq. (17) can be rewritten as $\boldsymbol{\beta} \sim e^{i\frac{\mathcal{S}}{\hbar}}$, where $\hbar$ is the reduced Planck constant, and $\mathcal{S}$ is by definition the generalized "action-integral":

$$\mathcal{S}(t_0, t_f) = \int_{t_0}^{t_f} dt\, \hbar \Omega_s(t), \tag{19}$$

The initial and final time instants are $t_0$ and $t_f$, respectively. The generalized action determines the coordinates of $\boldsymbol{\beta}$ in the great circle of the Bloch sphere ($\varphi = \frac{\mathcal{S}}{\hbar}$). The integrand of the action-integral controls how fast $\boldsymbol{\beta}$ sweeps the great circle of the sphere. When $\hbar\Omega_s$ is a constant, $\boldsymbol{\beta}$ sweeps every point of the great circle at the same rate, and hence there is no net translational motion, as seen in Sect. II.D. In contrast, when $\hbar\Omega_s$ varies with the particle position, it will reach a minimum for some vector $\boldsymbol{\beta}_{\min}$ of the time crystal cycle. The sweeping motion is slower when $\boldsymbol{\beta}$ is aligned with $\boldsymbol{\beta}_{\min}$, which means that $\boldsymbol{\beta}$ spends more time near $\boldsymbol{\beta}_{\min}$ than near the other directions of the great circle of the Bloch sphere. Consequently, the net translational motion of the massless particle must be towards the direction $\boldsymbol{\beta}_{\min}$. In other words, the particle moves towards the direction $\boldsymbol{\beta}$ corresponding to the minimum of the integrand of the action integral in a time-crystal cycle. It does so in a rather peculiar way: it probes the action integrand in every possible direction space in the plane perpendicular to the spin vector, and the net effect is a motion towards the direction that minimizes the integrand. For short time intervals with amplitude on the order of the time-crystal cycle, this mechanism is reminiscent of the Feynman's path integral of quantum mechanics where the time evolution of a quantum system is determined by the action evaluated along all possible paths.

In classical theory the action is determined by the Lagrangian ($\mathcal{L}$). This suggests that the spinning frequency must be controlled by a "Lagrangian": $\Omega_s \sim \mathcal{L}/\hbar$. This correspondence will be made precise in Sect. III.

The generalized action gives the winding number of $\boldsymbol{\beta}(t)$ in the Bloch sphere. The winding number (i.e., the phase $\varphi$) can change under a Lorentz boost. Thus, the generalized action is not a Lorentz invariant. Consider however a situation in which the initial and final points are identical: $\boldsymbol{\beta}(t_0) = \boldsymbol{\beta}(t_f)$. Then, the action is evidently a positive integer number $n$ in units of $2\pi\hbar$:

$\frac{\mathcal{S}}{2\pi\hbar} = \frac{1}{2\pi}\int_{t_0}^{t_f}\Omega_s(t)dt = n$. The integer $n$ gives the number of full loops (clock cycles) completed by $\boldsymbol{\beta}(t)$ around the great circle of the sphere in the time interval $t_0 \leq t \leq t_f$. In other words, the initial velocity is repeated exactly "$n$" times in the considered time interval: $\boldsymbol{\beta}(t_0) = \boldsymbol{\beta}(t_1) = .... = \boldsymbol{\beta}(t_n)$, with $t_n = t_f$. Clearly, the integer $n$ is frame independent because under a Lorentz boost the identity $\boldsymbol{\beta}(t_0) = \boldsymbol{\beta}(t_1) = .... = \boldsymbol{\beta}(t_n)$ becomes $\boldsymbol{\beta}'(t'_0) = \boldsymbol{\beta}'(t'_1) = .... = \boldsymbol{\beta}'(t'_n)$. Thus, it follows that for spacetime points separated by an integer number of "clock cycles" the generalized action is Lorentz invariant. As a first corollary, it follows that for a time interval with amplitude much larger than the clock cycle the action is "almost" Lorentz invariant (note that the action evaluated in different frames cannot differ by more than a Planck constant). The second corollary is that the action is exactly quantized in units of $2\pi\hbar$ for closed (periodic) orbits in the space domain.

In summary, the in-plane motion of a "time-crystal" particle (in the plane perpendicular to the spin vector) is governed by a generalized least action principle. For a planar or a quasi-planar motion, in each time-crystal cycle, the particle dynamically probes the action integral in every direction of space [Eq. (19)], and adjusts dynamically its motion towards the direction that minimizes the action.

### III. Center of mass frame

For a massive particle one can always find an inertial co-moving frame where the particle is instantaneously at rest. Evidently, such a concept cannot be directly applied to a relativistic massless particle as its speed is equal to *c* for any inertial observer.

Nevertheless, as a massless particle with nonzero acceleration follows a curved trajectory, one may wonder if it there is some special inertial reference frame (which may vary with time) wherein the particle follows to a good approximation a circular-type planar trajectory, with no net translational motion in space, analogous to the example worked out in the end of Sect. II.D. In what follows, I show that such a frame does indeed exist. It will be referred to as the "center of mass frame" or "co-moving frame", similar to the standard "rest frame" of massive particles. The velocity of the center of mass frame (normalized to the speed of light) in a generic (e.g., laboratory) frame is denoted by $\mathcal{V}(t)$.

### A. Definition and main properties

For a given spacetime point, the center of mass frame is defined as the unique inertial frame (primed coordinates) wherein the following conditions are satisfied:

$$\boldsymbol{\beta}' \cdot \mathbf{S}' = 0, \qquad |\dot{\boldsymbol{\beta}}'| = \frac{|\mathcal{L}'|}{\hbar}, \qquad \boldsymbol{\beta}' \cdot \frac{d\mathcal{V}'}{dt'} = 0. \qquad (20)$$

In the above, $\mathcal{L}'$ is the particle Lagrangian evaluated in the primed frame; it will be specified at a later point. For now, the Lagrangian should be regarded as some given function of the particle state. Moreover, $\mathcal{V}'$ represents the center of mass velocity measured in the primed inertial reference frame. By definition, the velocity $\mathcal{V}'$ vanishes at the spacetime point where the primed frame is coincident with the center of mass frame. Evidently, the center of mass frame typically changes with time and due to this reason $d\mathcal{V}'/dt'$ does not need to vanish at the same spacetime point. The motivation for the above construction is explained next.

The first condition, $\boldsymbol{\beta}' \cdot \mathbf{S}' = 0$, ensures that in the center of mass frame the torsion of the real-space trajectory vanishes, so that the motion is approximately planar and is contained in a plane perpendicular to the spin vector $\mathbf{S}'$, similar to a circular orbit. The second condition, establishes that the spin frequency in the co-moving frame ($\Omega'_s = |\dot{\boldsymbol{\beta}}'|$) is controlled by the Lagrangian of the system, in agreement with the generalized least action principle discussed in II.E Thus, the spin frequency is some pre-determined function of the particle state. It is worth pointing out that the relation $|\dot{\boldsymbol{\beta}}'| = \frac{|\mathcal{L}'|}{\hbar}$ is reminiscent of the quantum mechanics formula $E = \hbar \omega$ with $\omega \sim \Omega'_s$. Finally, the third condition establishes that in the co-moving frame the center of mass acceleration must be perpendicular to the wave velocity $\boldsymbol{\beta}$, analogous to the wave acceleration which satisfies

-14-

$\boldsymbol{\beta} \cdot \dot{\boldsymbol{\beta}} = 0$. It will be seen in section III.B that the third condition is essential to guarantee that the time derivative of the center of mass momentum transforms as a relativistic force.

For completeness, it is demonstrated in the supplementary information [13, Sect. D] that for a given trajectory there is indeed a unique inertial frame wherein the conditions (20) are satisfied for each spacetime point. The relative velocity $\mathcal{V}$ of the center of mass frame with respect to a fixed laboratory (unprimed) frame can be written explicitly in terms of the kinematic parameters of the trajectory ($\boldsymbol{\beta}, \dot{\boldsymbol{\beta}}, \ddot{\boldsymbol{\beta}}$) and of the Lagrangian function $\mathcal{L}$ [13, Eq. (D11)]. Crucially, it turns out that $c\mathcal{V}$ transforms as a relativistic velocity under a Lorentz boost and is always less than $c$: $|\mathcal{V}| < 1$. Thus, it is possible to associate *two* velocities with a time-crystal particle: the velocity of the real-space trajectory $\boldsymbol{\beta}$ and the velocity of the center of mass $\mathcal{V}$. Evidently, the center of mass frame changes continuously with time, and thereby $\mathcal{V}$ is generally a function of time that depends on the particular trajectory of the time-crystal particle.

In summary, a time-crystal particle is characterized by a Lorentz co-variant center of mass frame defined by the conditions (20). In the center of mass frame, there is no net translational motion, only a rotation about some curvature center.

### B. Center of mass energy-momentum 4-vector

The center of mass velocity $\mathcal{V}$ transforms as a relativistic velocity under a Lorentz boost [13]. Due to this reason, it is possible to construct a energy-momentum 4-vector associated with $\mathcal{V}$. Since $|\mathcal{V}| < 1$, it is logical to take the energy-momentum 4-vector $(E, \boldsymbol{\pi})$ of the form:

$$\boldsymbol{\pi} = m_e \gamma_\mathcal{V} c \mathcal{V}, \qquad E = \gamma_\mathcal{V} m_e c^2. \qquad (21)$$

Here, $\gamma_\mathcal{V} = 1/\sqrt{1 - \mathcal{V} \cdot \mathcal{V}}$ is center of mass Lorentz factor and $m_e$ is a coupling constant that will be identified with the usual rest mass of the classical particle. The center of mass velocity can be expressed in terms of the massive (center of mass) momentum $\boldsymbol{\pi}$, in the usual way:

$$\mathcal{V} = \frac{\boldsymbol{\pi}}{\sqrt{\boldsymbol{\pi} \cdot \boldsymbol{\pi} + (m_e c)^2}}. \qquad (22)$$

It is underlined that $(E, \boldsymbol{\pi})$ transforms as a 4-vector simply because $\mathcal{V}$ transforms as a relativistic velocity. It will be seen in Sect. IV, that $(E, \boldsymbol{\pi})$ can be identified with the classical energy-momentum 4-vector of the particle.

Hence, there are three different 4-vectors associated with a time-crystal particle: the spin 4-vector $(\mathbf{S}\cdot\boldsymbol{\beta}, \mathbf{S})$, the massless (wave) energy-momentum 4-vector $(\mathcal{E}, \mathcal{P})$, and the massive energy-momentum 4-vector $(E, \boldsymbol{\pi})$. Note that $\mathcal{P}$ is a massless type momentum ($|\mathcal{E}|^2 - c^2|\mathcal{P}|^2 = 0$), the spin vector is a massive time-like 4-vector ($(\mathbf{S}\cdot\boldsymbol{\beta})^2 - \mathbf{S}\cdot\mathbf{S} = -1$), and $\boldsymbol{\pi}$ is a massive space-like momentum ($|E|^2 - c^2|\boldsymbol{\pi}|^2 = m_e^2 c^4$).

## *C. Spin frequency*

As discussed previously, the spinning frequency in the co-moving frame must be controlled by the Lagrangian of the particle. In classical theory, the Lagrangian of a relativistic charged classical particle with mass $m$ and charge $q_e$ is $\mathcal{L} = -mc^2/\gamma + q_e(\mathbf{v}\cdot\mathbf{A} - \varphi)$ [14]. Here, $(\varphi/c, \mathbf{A})$ is the electromagnetic 4-potential and $\gamma$ is the Lorentz factor. An obvious analogy with classical theory, suggest that in the co-moving frame $\Omega'_s = |\mathcal{L}'|/\hbar = \left[m_e c^2 + q_e(\varphi' - c\boldsymbol{\beta}'\cdot\mathbf{A}')\right]/\hbar$. On the other hand, as $|\dot{\boldsymbol{\beta}}|\mathcal{E}^2$ is a Lorentz scalar, it is compelling to write the spin frequency as in Eq. (7). This can be ensured by taking $(E_1, \mathbf{p}_1)$ as linear combination of the massive energy-momentum and of the electromagnetic 4-potential, $(E_1, \mathbf{p}_1) \to (E_{can}, \mathbf{p}_{can})$ and $(E_2, \mathbf{p}_2)$ identical to the massive energy-momentum $(E_2, \mathbf{p}_2) \to (E, \boldsymbol{\pi})$. This leads to the manifestly co-variant formula for the spin frequency:

$$|\dot{\boldsymbol{\beta}}| = \frac{1}{\hbar m_e c^2}(E_{can} - c\mathbf{p}_{can}\cdot\boldsymbol{\beta})(E - c\boldsymbol{\pi}\cdot\boldsymbol{\beta}). \tag{23}$$

Here, $(E_{can}, \mathbf{p}_{can})$ is the classical canonical energy-momentum given by:

$$E_{can} = E + \theta_e q_e \varphi, \qquad \mathbf{p}_{can} = \boldsymbol{\pi} + \theta_e q_e \mathbf{A}. \tag{24}$$

The parameter $\theta_e$ is a renormalization parameter that will be fixed later (this parameter can always be set identical to one with a suitable charge renormalization; for simplicity, in this article the charge and mass parameters are identified with the experimental values and due to this reason $\theta_e \neq 1$). Note that in the co-moving frame $E = m_e c^2$, $\boldsymbol{\pi} = 0$, and thereby the spin angular frequency reduces to $\Omega'_s = \left[ m_e c^2 + \theta_e q_e (\varphi' - c\boldsymbol{\beta}' \cdot \mathbf{A}') \right]/\hbar$, which is consistent with the Lagrangian form (with the renormalization parameter $\theta_e$ included in the Lagrangian).

The Lagrangian is given by the (Lorentz-covariant) formula $\mathcal{L} = -\left( E_{\text{can}} - c\mathbf{p}_{\text{can}} \cdot \boldsymbol{\beta} \right)$, which ensures that under a Lorentz boost it transforms as $\mathcal{L}' = \mathcal{L}/(\gamma D)$ (compare with the acceleration in Eq. (5)). This property guarantees that the standard action integral $\int dt \mathcal{L}$ is always a relativistic invariant.

### D. Dynamics of the center of mass

It is shown in the supplementary information [13, Sect. E] that the condition $\boldsymbol{\beta}' \cdot d\boldsymbol{\mathcal{V}}'/dt' = 0$ in Eq. (20) implies that the time derivative of the kinetic momentum $d\boldsymbol{\pi}/dt$ in a generic inertial frame is constrained by:

$$(\boldsymbol{\beta} - \boldsymbol{\mathcal{V}}) \cdot \frac{d\boldsymbol{\pi}}{dt} = 0 . \tag{25}$$

Moreover, it is proven that due to $\boldsymbol{\beta}' \cdot d\boldsymbol{\mathcal{V}}'/dt' = 0$ the time derivative of the kinetic momentum $d\boldsymbol{\pi}/dt$ transforms as a relativistic force under a Lorentz boost. Thus, the third constraint in the center of mass definition is essential to guarantee that $d\boldsymbol{\pi}/dt$ can be regarded as a relativistic force.

Let us denote $d\boldsymbol{\pi}/dt$ as $\mathcal{F}$. For a charged particle, let us say a "time-crystal electron", it is tempting to impose that $\mathcal{F}$ is governed by the standard Lorentz force, so that $\mathcal{F} = q_e (\mathbf{E} + c\boldsymbol{\mathcal{V}} \times \mathbf{B})$ with $q_e = -e$ the elementary charge and $\mathbf{E}$ and $\mathbf{B}$ the electric and magnetic fields, respectively. However, this is not feasible because the force $\mathcal{F}$ must satisfy $(\mathcal{V}-\boldsymbol{\beta})\cdot\mathcal{F}=0$ at all time instants [Eq. (25)], and evidently the Lorentz force is incompatible with this requirement.

In the $\mathcal{V}=0$ frame, the vectors $\boldsymbol{\beta}\times\mathbf{S}, \boldsymbol{\beta}, \mathbf{S}$ form an orthonormal right-handed basis. Thus, the force in the co-moving frame, $\mathcal{F}_{co}$, can always be written as a linear combination of the vectors $\boldsymbol{\beta}\times\mathbf{S}$ and $\mathbf{S}$: $\mathcal{F}_{co}=c_1\boldsymbol{\beta}\times\mathbf{S}+c_2\mathbf{S}$. The first term ($c_1\boldsymbol{\beta}\times\mathbf{S}$) gives the force projection onto the plane of motion of the time-crystal particle, and the second term ($c_2\mathbf{S}$) determines the force along the direction perpendicular to that plane. Taking into account how the velocity and spin vector transform under a Lorentz boost one may show that the force in a generic frame is given by the following (Lorentz-covariant) expression (for convenience I use $c_1 \to L_{in}/R_C$ and $c_2 \to L_{out}/R_C$):

$$\mathcal{F}=\gamma_\mathcal{V}\left[\frac{L_{in}}{R_C}(\boldsymbol{\beta}-\mathcal{V})\times\tilde{\mathbf{S}}+\frac{L_{out}}{R_C}(1-\mathcal{V}\cdot\boldsymbol{\beta})\mathbf{S}\right], \tag{26}$$

where $\tilde{\mathbf{S}}=\mathbf{S}-\mathcal{V}(\mathbf{S}\cdot\boldsymbol{\beta})$, $\gamma_\mathcal{V}=1/\sqrt{1-\mathcal{V}\cdot\mathcal{V}}$ and $R_C$ is a parameter with dimensions of length taken equal to $R_C\equiv\frac{\hbar}{m_e c}$ (for an electron $R_C$ is the reduced Compton wavelength given by roughly 1/137 times the Bohr radius). In the above, $L_{in}, L_{out}$ are functions with units of energy that transform as Lorentz scalars, and control the in-plane and out-of-plane force components, respectively. A detailed characterization of these functions will be presented in Sect. V. It may be checked that $\mathcal{F}$ given by the above formula satisfies $\mathcal{F}\cdot(\mathcal{V}-\boldsymbol{\beta})=0$.

The standard Lorentz force is even under a time-reversal (T) transformation and odd under a parity (P) transformation. In order that $\mathcal{F}$ given by Eq. (26) has the same property it is necessary that $L_{in}$ ($L_{out}$) is even (odd) under both the P and T transformations.

## E. Dynamics of the spin vector

Next, it is shown that the dynamics of the spin vector is controlled by the center of mass acceleration. The spin vector must satisfy two purely kinematic constraints. Namely, since $(\mathbf{S}\cdot\boldsymbol{\beta}, \mathbf{S})$ is a 4-vector it is necessary that:

$$(\mathbf{S}\cdot\boldsymbol{\beta})^2 - \mathbf{S}\cdot\mathbf{S} = -1, \qquad \text{and} \qquad \mathcal{V}\cdot\mathbf{S} = \mathbf{S}\cdot\boldsymbol{\beta}. \tag{27}$$

The first condition is a repetition of Eq. (11), while the second condition follows from the center of mass definition ($\boldsymbol{\beta}'\cdot\mathbf{S}' = 0$) [Eq. (20)] and from Eq. (10). The time evolution of the spin vector must ensure that the constraints (27) are satisfied at any time instant.

As previously discussed, the spin vector is proportional to the binormal of the velocity trajectory, $\mathbf{S} \sim \hat{\mathbf{b}}$ [Eq. (9)]. From the Frenet-Serret formulas [12], the derivative in time of the binormal is parallel to the normal vector: $d\hat{\mathbf{b}}/dt \sim \hat{\mathbf{n}} = \hat{\mathbf{b}}\times\hat{\mathbf{t}}$. The tangent vector is proportional to the acceleration $\dot{\boldsymbol{\beta}}$ (see Sect. II.C). Hence, from Eq. (8) and $\mathbf{S} \sim \hat{\mathbf{b}}$, one sees that $d\mathbf{S}/dt$ must be a linear combination of $\mathbf{S}$ and $\mathbf{S}\times(\mathbf{S}\times\boldsymbol{\beta})$:

$$\frac{d\mathbf{S}}{dt} = \alpha_1 \mathbf{S} + \alpha_2 (\boldsymbol{\beta}\times\mathbf{S})\times\mathbf{S}, \tag{28}$$

where $\alpha_i$ are some unknown coefficients. Differentiating $(\mathbf{S}\cdot\boldsymbol{\beta})^2 - \mathbf{S}\cdot\mathbf{S} = -1$ with respect to time one gets $\frac{d\mathbf{S}}{dt}\cdot\boldsymbol{\beta}(\mathbf{S}\cdot\boldsymbol{\beta}) - \mathbf{S}\cdot\frac{d\mathbf{S}}{dt} = 0$, where it was taken into account that the acceleration is always perpendicular to the spin vector $\dot{\boldsymbol{\beta}}\cdot\mathbf{S} = 0$ [Eq. (8)]. The derived relation can be satisfied only if $\alpha_1 + \alpha_2 \mathbf{S}\cdot\boldsymbol{\beta} = 0$. Substituting $\alpha_1 = -\alpha_2 \mathbf{S}\cdot\boldsymbol{\beta}$ into Eq. (28) and simplifying one finds that $\frac{d\mathbf{S}}{dt} = \alpha\boldsymbol{\beta}$ with $\alpha$ some unknown coefficient. In other words, the time derivative of the spin vector must be proportional to the velocity of the time-crystal particle.

To make further progress, I differentiate the second kinematic constraint, $\mathcal{V}\cdot\mathbf{S} = \mathbf{S}\cdot\boldsymbol{\beta}$, with respect to time. This yields the relation $\frac{d\mathbf{S}}{dt}\cdot(\boldsymbol{\beta}-\mathcal{V}) = \frac{d\mathcal{V}}{dt}\cdot\mathbf{S}$. Using now $\frac{d\mathbf{S}}{dt} = \alpha\boldsymbol{\beta}$ in the previous formula, one finds that $\alpha(1-\boldsymbol{\beta}\cdot\mathcal{V}) = \frac{d\mathcal{V}}{dt}\cdot\mathbf{S}$, so that

$$\frac{d\mathbf{S}}{dt} = \frac{\frac{d\mathcal{V}}{dt}\cdot\mathbf{S}}{(1-\boldsymbol{\beta}\cdot\mathcal{V})}\boldsymbol{\beta}. \tag{29}$$

An immediate consequence of Eq. (29) is that the spin vector is conserved when the center of mass velocity is constant, or more generally when $\mathbf{S}\cdot d\mathcal{V}/dt = 0$, e.g., for planar trajectories.

Moreover, using Eq. (22) it can be readily shown that $\frac{d\mathcal{V}}{dt} = \frac{1}{m_e c \gamma_\mathcal{V}}(\mathbf{1}-\mathcal{V}\otimes\mathcal{V})\cdot\frac{d\boldsymbol{\pi}}{dt}$. Substituting this formula and Eq. (26) into Eq. (29), one finds with the help of (27) the remarkably simple result

$$\frac{d\mathbf{S}}{dt} = \frac{L_{\text{out}}}{\hbar}\boldsymbol{\beta}. \tag{30}$$

Hence, the dynamics of the spin vector is fully determined by the time derivative of the center of mass velocity, or equivalently by the Lorentz scalar $L_{\text{out}}$. The derived law is purely kinematic and applies to any inertial frame, and thereby it is Lorentz covariant. Furthermore, it can be checked that similar to $d\boldsymbol{\pi}/dt$ the time derivative of the spin vector transforms as a relativistic force under a Lorentz boost.

To conclude, I note that from $\dot{\boldsymbol{\beta}} = \Omega_s \mathbf{S}\times\boldsymbol{\beta}$ [Eq. (8)] it follows that:

$$\begin{aligned}\dot{\boldsymbol{\beta}}\times\ddot{\boldsymbol{\beta}} &= (\Omega_s\mathbf{S}\times\boldsymbol{\beta})\times\left[\Omega_s\mathbf{S}\times\dot{\boldsymbol{\beta}} + \Omega_s\dot{\mathbf{S}}\times\boldsymbol{\beta} + \frac{d\Omega_s}{dt}\mathbf{S}\times\boldsymbol{\beta}\right] \\ &= \Omega_s^2(\mathbf{S}\times\boldsymbol{\beta})\times(\mathbf{S}\times\dot{\boldsymbol{\beta}}) = \Omega_s|\dot{\boldsymbol{\beta}}|^2\mathbf{S}\end{aligned} \tag{31}$$

where $\frac{d\mathbf{S}}{dt} \sim \boldsymbol{\beta}$ [Eq. (29)] was used in the second identity. This confirms that Eq. (29) really describes the spin vector given by Eq. (9), i.e., $\mathbf{S}$ is indeed proportional to the binormal of the velocity trajectory in the Bloch sphere.

## IV. "Pilot-wave" mechanical model

### A. Mass as an emergent property

As seen in Sect. III, it is possible to assign a center of mass velocity $\mathcal{V} = \mathcal{V}(t)$ to a massless particle satisfying the postulates P1 and P2, as well as a massive energy-momentum 4-vector $(E, \boldsymbol{\pi})$, with $d\boldsymbol{\pi}/dt$ transforming as a force. Thus, the next logical step is to integrate the velocity $\mathcal{V}(t)$, and in this manner obtain a center of mass trajectory $\mathbf{r}_{CM} = \mathbf{r}_{CM}(t)$. Evidently, the integration of $\mathcal{V}(t)$ is defined up to an arbitrary integration constant. However, here I want to attribute some actual physical reality to the "center of mass" that removes such arbitrariness. In fact, the postulates P1 and P2 lead naturally to the idea that the time-crystal particle is formed by two components: the massless component described by $\mathbf{r}_0 = \mathbf{r}_0(t)$, and the center of mass described by $\mathbf{r}_{CM} = \mathbf{r}_{CM}(t)$. The two components are not different particles, but rather different aspects of the same physical object: the "time-crystal".

There is a difficulty: the equation $\frac{d\mathbf{r}_{CM}}{dt}(t) = c\mathcal{V}(t)$ is not Lorentz covariant. Furthermore, since $\mathcal{V}(t)$ is a vector determined by the parameters of the trajectory $\mathbf{r}_0 = \mathbf{r}_0(t)$, an equation of the type $\frac{d\mathbf{r}_{CM}}{dt}(t) = c\mathcal{V}(t)$ would imply that a change in trajectory $\mathbf{r}_0 = \mathbf{r}_0(t)$ would affect instantaneously the $\mathbf{r}_{CM}$ trajectory. This would require a superluminal mechanism.

Thus, to construct a Lorentz covariant theory one must proceed more carefully. The simplest solution is to impose that [13, Sect. G]:

$$\frac{d\mathbf{r}_{CM}}{dt}(t) = c\boldsymbol{\mathcal{V}}(t_d), \qquad \text{with} \qquad t = t_d + \frac{1}{c}|\mathbf{r}_{CM}(t) - \mathbf{r}_0(t_d)|. \tag{32}$$

In this manner, $\boldsymbol{\mathcal{V}}_{CM} \equiv d\mathbf{r}_{CM}/dt$ is determined by a time-delayed version of the velocity of the electron center of mass. Note that $t > t_d$ and thereby the above equation is causal.

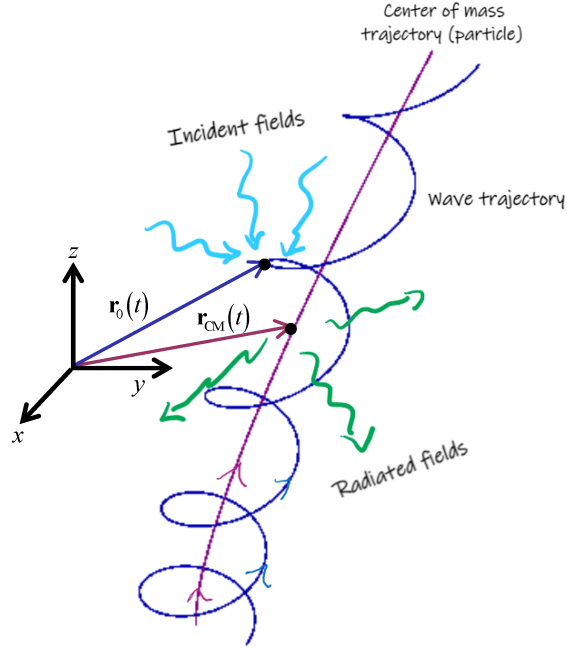

**Fig. 2** Sketch of the trajectories of the two components of a time-crystal particle: the wave-component $\mathbf{r}_0(t)$ (helical-type path in blue) and the particle-component $\mathbf{r}_{CM}(t)$ (center of mass trajectory, in purple). A charged time-crystal particle interacts with incident (e.g., external) fields through the wave-component $\mathbf{r}_0(t)$. The field radiated by the particle emerges from the center of mass position. Since the incoming and outgoing waves are coupled to different trajectories, the self-field interactions do not lead to any type of singularities.

With the previous step one has come full circle: starting from a massless particle, one arrives naturally to the conclusion that it must have as well a massive component, consistent with classical theory. The particle's mass is an emergent property as it follows from the fact that $|\boldsymbol{\mathcal{V}}| < 1$. Different from classical physics, the time-crystal particle also has spin and a "massless" energy-momentum 4-vector $(\mathcal{E}, \mathcal{P})$. Moreover, rather peculiarly, a single particle is characterized by two space trajectories [Fig. 2]. This property establishes a peculiar form of particle-wave duality.

It should be noted that because of the time delay, the velocity $\mathcal{V}_{CM}$ is typically nonzero in the "center of mass frame", i.e., $\mathcal{V}_{CM}(t) \neq 0$ in the inertial frame wherein $\mathcal{V}(t) = 0$. In other words, the frame instantaneously co-moving with the $\mathbf{r}_{CM}$-trajectory should not be confused with the "center of mass frame" of the $\mathbf{r}_0$-trajectory introduced earlier. The energy-momentum 4-vector associated with the $\mathbf{r}_{CM}$-trajectory is further discussed in the supplementary information [13, Sect. F].

## B. Pilot-wave analogy

The two-component model for the time-crystal particle is reminiscent of the pilot-wave theory for quantum particles first proposed by L. de Broglie and later rediscovered by D. Bohm [10-11]. The de Broglie-Bohm theory relies on the idea that a quantum particle consists of a "point particle" that is guided by an associated "matter wave" (the pilot wave). The particle follows a trajectory controlled by the pilot-wave. Different from the most common (Copenhagen) interpretation of quantum mechanics, the de Broglie–Bohm theory is deterministic as it enables a precise and continuous description of all physical processes, independent of the observer and of the measurement process.

Curiously, in the same manner as the pilot wave guides the particle in the de Broglie–Bohm framework, in the model introduced here the massless component of the particle governs the trajectory $\mathbf{r}_{CM}$ through Eq. (32). In fact, the velocity $\mathcal{V}$ is completely determined by the trajectory $\mathbf{r}_0$ [13, Eq. (D11)], and thereby the trajectory $\mathbf{r}_{CM}$ is fully controlled by the properties of the massless component. Due to this reason, the component associated with $\mathbf{r}_{CM}$ will be referred to as the "particle-component" (or center-of-mass component) of the time-crystal, whereas the component associated with $\mathbf{r}_0$ will be referred to as the "wave-component" (or massless component). The wave-component whirls around the particle-component at the speed of light and effectively probes the near-by space (Fig. 2).

Moreover, analogous to de Broglie–Bohm theory, the time-crystal model has nonlocal features. In fact, as soon as a single particle has two components, each with its own individual trajectory, it may be influenced simultaneously by what happens at two distinct points in space. It is worth pointing out that the Bell's theorem states that the predictions of quantum mechanics are incompatible with any local theory of hidden variables [15]. Thereby, any deterministic model that can capture some of the weirdness of quantum physics is expected to have nonlocal features.

To conclude this sub-section, it is worth alluding to a fascinating example of a pilot-wave mechanical system of a totally different kind that was introduced and experimentally studied by Couder and Fort in Refs. [16]-[18]. The Couder-Fort system consists of a bouncing droplet (particle) that travels in a vertically vibrated membrane. The particle is guided by the "pilot-wave" generated by its resonant interaction with the membrane. Interestingly, Couder and Fort were able to observe diffraction and interference phenomena in a "bouncing droplet" version of the double-slit experiment [16], observe the quantization of classical orbits [17], and observe a macroscopic analogue of the Zeeman splitting [18]. Even though some of their findings related to the double-slit experiment have been disputed by other researchers [19], their discoveries led to a renewed interest in hidden variable models of the quantum world [20].

### C. Coupling with the electromagnetic field

Next, I briefly outline how a charged time-crystal particle may be coupled to the electromagnetic field.

A charged time-crystal particle must be coupled to the Maxwell's equations through the center of mass coordinates ($\mathbf{r}_{CM}$) (see Appendix A for a detailed discussion). This means that the radiated fields emerge from the center of mass (Fig. 2). On the other hand, the dynamics of the electron is controlled by the generalized Lorentz force [Eq. (26)], which will depend on the electromagnetic fields evaluated at the wave coordinates $(\mathbf{r}_0, t)$. Importantly, this property implies that the singular point of the emitted fields $(\mathbf{r}_{CM}, t)$ is not coincident with the field point $(\mathbf{r}_0, t)$ that controls the dynamics of the time-crystal particle (see Fig. 2). In other words, the interaction of the time-crystal electron with the Maxwell field is nonlocal because the incoming and outgoing electromagnetic waves need to be coupled to different components of the time-crystal (see Fig. 2). Thus, quite remarkably, the time-crystal model and its nonlocal features may naturally offer a solution to the problem of infinities of the self-energy and self-force that plague both classical and quantum electrodynamics [21, 22-24].

Moreover, the coupling with the Maxwell field suggests that the wave energy $\mathcal{E}$ is related to the electromagnetic interaction of the time-crystal with itself. Specifically, the wave energy in the co-moving frame is roughly identified with the self Coulomb potential: $\mathcal{E}_{co} \sim q_e^2 / \left( 4\pi\varepsilon_0 \left| \mathbf{r}_0(t_d) - \mathbf{r}_{CM}(t) \right| \right)$. It is out of the intended scope of the article to analyze the self-field effects or develop fully the coupling with the Maxwell field. For simplicity, in the rest of this article the wave energy will be taken as a constant $\mathcal{E}_{co} = \mathcal{E}_{self,EM} = const.$ in the co-moving frame.

### D. Spin angular momentum

Due to the postulate P2 the pilot-wave trajectory is necessarily curved. Let us show that there is a conservation law associated with the spinning motion. Specifically, from $\frac{d}{dt}(\mathbf{r}_0 \cdot \boldsymbol{\beta}) = \mathbf{r}_0 \cdot \dot{\boldsymbol{\beta}} + c\boldsymbol{\beta} \cdot \boldsymbol{\beta}$ and Eq. (8), it readily follows that $\frac{d}{dt}\left( \mathbf{r}_0 \cdot \frac{\boldsymbol{\beta}}{c} \right) + \mathbf{S} \cdot \left( \mathbf{r}_0 \times \frac{\Omega_s \boldsymbol{\beta}}{c} \right) = 1$. This means that for orbits that are periodic in time (i.e., closed orbits in some particular inertial frame), the time average of $\mathbf{S} \cdot \left( \mathbf{r}_0 \times \frac{\Omega_s \boldsymbol{\beta}}{c} \right)$ in a full orbit cycle is a constant ($T$ is the period of the orbit):

$$\frac{1}{T} \int_0^T \frac{\Omega_s}{c^2} \mathbf{S} \cdot (\mathbf{r}_0 \times c\boldsymbol{\beta}) \, dt = 1, \qquad \text{(stationary orbits).} \qquad (33)$$

It is underscored that this result holds true independent of the origin of space and in a frame where the orbit is stationary. The result is also independent of the specific variation in time of $\Omega_s$ and $\mathbf{S}$. Let us now suppose that in the relevant frame the center of mass velocity is much smaller than the speed of light. Then, using the approximation $\Omega_s \approx \Omega_{s,co} \approx m_e c^2/\hbar$ it follows that $\left\langle \frac{m_e}{2} \mathbf{S} \cdot (\mathbf{r}_0 \times \boldsymbol{\beta} c) \right\rangle \approx \frac{\hbar}{2}$. The quantity $L_{spin} = \left\langle \frac{m_e}{2} \mathbf{S} \cdot (\mathbf{r}_0 \times \boldsymbol{\beta} c) \right\rangle$ may be regarded as the projection of an angular momentum onto the spin vector. Similar to a spin ½ quantum particle, for stationary orbits $L_{spin}$ has the universal value $L_{spin} \approx \hbar/2$.

### E. Partial summary of the model

Here, I recapitulate the main results obtained so far. Starting from the postulates P1 and P2 and from the center of mass definition [Eq. (20)], it was proven that the time-crystal particle is characterized by massless (wave) and massive (particle) components, with trajectories controlled by:

$$\frac{d\mathbf{r}_0}{dt} = c\boldsymbol{\beta}, \qquad \frac{d\mathbf{r}_{CM}}{dt} = c\boldsymbol{\mathcal{V}}_{CM}, \tag{34a}$$

where $\boldsymbol{\mathcal{V}}_{CM}(t) = \boldsymbol{\mathcal{V}}(t_d)$ and $t = t_d + \frac{1}{c}|\mathbf{r}_{CM}(t) - \mathbf{r}_0(t_d)|$. There are three 4-vectors, $(\mathcal{E}, \mathcal{P})$, $(E, \boldsymbol{\pi})$, and $(\mathbf{S} \cdot \boldsymbol{\beta}, \mathbf{S})$, associated with the time-crystal particle. The center of mass definition ensures that the trajectory in the co-moving frame is approximately planar and has a curvature controlled by the classical Lagrangian. In addition, it ensures that the time-derivative of the kinetic momentum, $d\boldsymbol{\pi}/dt$, transforms as relativistic force in both the $\mathbf{r}_0$ and $\mathbf{r}_{CM}$ trajectories.

The dynamics of the wave velocity and spin vector are ruled by

$$\frac{d\boldsymbol{\beta}}{dt} = \Omega_s \mathbf{S} \times \boldsymbol{\beta}, \quad \text{with} \quad \Omega_s = \frac{1}{\hbar m_e c^2}(E_{can} - c\mathbf{p}_{can} \cdot \boldsymbol{\beta})(E - c\boldsymbol{\pi} \cdot \boldsymbol{\beta}), \tag{34b}$$

$$\frac{d\mathbf{S}}{dt} = \frac{L_{out}}{\hbar}\boldsymbol{\beta}, \tag{34c}$$

where $(E_{can}, \mathbf{p}_{can})$ is the canonical energy-momentum [Eq. (24)]. The wave dynamics is controlled by a generalized least action principle. Finally, the dynamics of the massive momentum is governed by:

$$\frac{d\boldsymbol{\pi}}{dt} = \mathcal{F} = \gamma_\mathcal{V} \left[ \frac{L_{in}}{R_C}(\boldsymbol{\beta} - \mathcal{V}) \times \tilde{\mathbf{S}} + \frac{L_{out}}{R_C}(1 - \mathcal{V} \cdot \boldsymbol{\beta})\mathbf{S} \right]. \tag{34d}$$

The time evolution of the time-crystal state is fully determined by the functions (Lorentz scalars) $L_{in}, L_{out}$. Explicit formulas for $L_{in}, L_{out}$ will be given in Sect. V. It can be verified that the system of equations (34) is Lorentz covariant.

For charged particles, the functions $L_{in}, L_{out}$ depend on the external electromagnetic fields evaluated at the coordinates of the wave-component $(\mathbf{r}_0, t)$, and thus the force $\mathcal{F}$ has the same property. As a consequence, the dynamics of the center of mass $(\mathbf{r}_{CM}, t)$ will be controlled by the optical field at the point $(\mathbf{r}_0, t)$, which results in a theory with nonlocal features.

### F. Free-particle with $\Omega_s|_{co\text{-}moving} = const.$

Let us now analyze in detail the trajectory of a "free-particle", i.e., by definition a time-crystal free from any interactions ($L_{in} = 0 = L_{out}$), including the self electromagnetic interactions ($(\varphi, \mathbf{A}) = 0$). For a free-particle, the spin frequency in the co-moving frame is a constant of motion: $\Omega_{co} = const. \equiv \Omega_0 = m_e c^2 / \hbar$ [Eq. (23)].

From Eqs. (34c) and (34d), one sees that when $L_{in} = 0 = L_{out}$ the spin vector $\mathbf{S}$ and the massive momentum $\boldsymbol{\pi}$ are also constants of motion. Furthermore, from Eq. (22) the center-of-mass velocity is also time-invariant ($\mathcal{V} = const.$) so that the center of mass trajectory is a straight line ($\mathbf{r}_{CM}(t) = \mathbf{r}_{CM}(0) + c\mathcal{V}t$).

Since $(\mathbf{S} \cdot \boldsymbol{\beta})^2 - \mathbf{S} \cdot \mathbf{S} = -1$ the result $\mathbf{S} = const.$ implies that $\mathbf{S} \cdot \boldsymbol{\beta} = const. \equiv |\mathbf{S}|\cos\theta$, with $\theta$ the angle introduced in Sect. II.C determined by the wave velocity ($\boldsymbol{\beta}$) and the spin vector in the Bloch sphere. For a free-particle $\theta = \theta_0$ is necessarily time invariant, and thereby the velocity trajectory in the Bloch sphere is constrained to be in a circle perpendicular to the spin vector (see Fig. 1). The radius of the circle is evidently $\sin\theta_0$, and its inverse gives the curvature $K$ of the trajectory of $\boldsymbol{\beta}$ in the Bloch sphere. This result is in agreement with $K = |\mathbf{S}| = 1/\sin\theta_0$ [Eq. (12)]. Note that $\boldsymbol{\beta}$ cannot be parallel to the spin vector because this would give $|\mathbf{S}| = \infty$.

For $\theta_0 \neq 90º$ there is necessarily a net motion of the time-crystal particle along the direction of the spin vector, and hence the trajectory of the particle must be unbounded in space. This property is in agreement with $\mathcal{V} \cdot \mathbf{S} = \mathbf{S} \cdot \boldsymbol{\beta}$ [Eq. (27)], which shows that for $\theta_0 \neq 90º$ the center of mass velocity cannot vanish. Furthermore, for $\theta_0 \neq 90º$ the trajectory of the free particle cannot be planar because it has a nontrivial torsion [Eq. (15)]. Indeed, as already noted in Sect. II.C, the trajectory may be contained in a plane ($\tau_R = 0$) only when the time component of the spin 4-vector vanishes ($\mathbf{S} \cdot \boldsymbol{\beta} = 0$), i.e., when $\theta_0 = 90º$.

Even though the wave velocity whirls around $\mathbf{S}$, the net velocity of the time-crystal particle ($\mathcal{V}$) does not need to be parallel to $\mathbf{S}$. In fact, the direction of the center of mass motion depends also on the rate at which the $\boldsymbol{\beta}$ sweeps the circle $\theta = \theta_0$, which does not need to be uniform. Due to this reason, it is possible to have a net translational motion in the plane perpendicular to the spin vector towards the direction determined by the minimum of $|\dot{\boldsymbol{\beta}}|$ in the circle $\theta = \theta_0$. Using $(\varphi, \mathbf{A}) = 0$ in Eq. (23) one gets $\Omega_s = \frac{1}{\hbar m_e c^2}(E - c\boldsymbol{\pi} \cdot \boldsymbol{\beta})^2$ and from the definition of the massive energy-momentum it follows that:

$$|\dot{\boldsymbol{\beta}}| = \Omega_s = \Omega_0 \frac{(1 - \mathcal{V} \cdot \boldsymbol{\beta})^2}{1 - \mathcal{V} \cdot \mathcal{V}}, \qquad \text{with} \quad \Omega_0 = \frac{m_e c^2}{\hbar}, \tag{35}$$

Clearly, the acceleration can vary in each time crystal cycle as the term $\mathcal{V} \cdot \boldsymbol{\beta}$ is a function of time. This idea is further developed below.

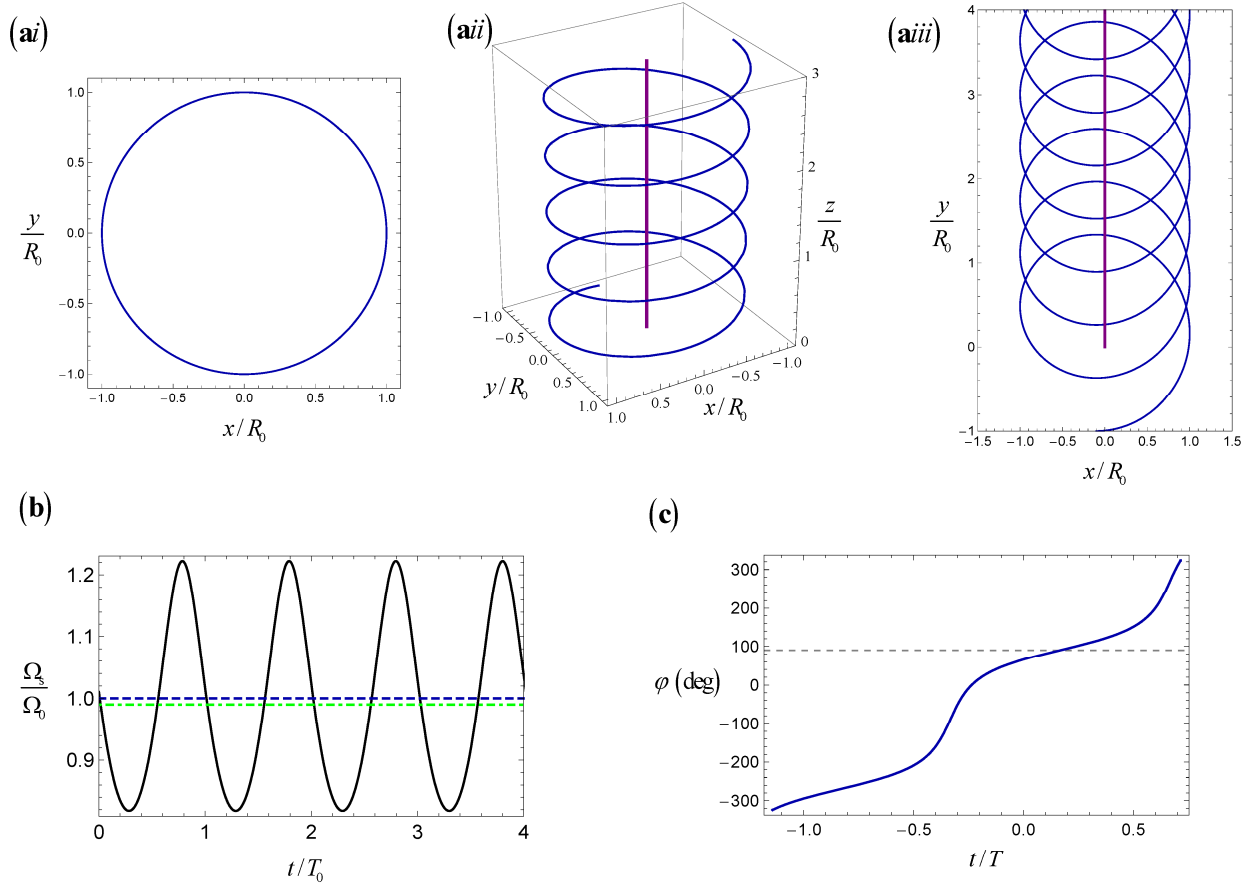

**Fig. 3 (a)** Free particle trajectories with $\Omega_s|_{\text{co-moving}} = \Omega_0 = const.$. The spin vector **S** is directed along the $+z$-direction. Blue lines: trajectory of the wave-component $\mathbf{r}_0$. Purple lines: trajectory of the particle-component $\mathbf{r}_{CM}$. i) $\mathcal{V} = 0$. ii) Longitudinal motion with $\mathcal{V} = 0.1\hat{\mathbf{z}}$. iii) Transverse motion with $\mathcal{V} = 0.1\hat{\mathbf{y}}$. **(b)** Normalized spin frequency $\Omega_s$ for cases (ai) (blue line), (aii) (green line) and (aiii) (black line). **(c)** Representation of $\varphi(t)$ as a function of time for a transverse motion with $\mathcal{V} = 0.5\hat{\mathbf{y}} \sim 0.5i$. The angle $\varphi(t)$ varies slowly in time when $\mathcal{V}$ and $\boldsymbol{\beta} \sim \exp(i\varphi)$ are nearly parallel, i.e., when $\varphi = 90º$ (horizontal gray line).

Evidently, for a free-particle it is always possible to switch to an inertial frame coincident with the center of mass frame where $\mathcal{V} = 0$. In the center of mass frame the trajectory is planar ($\boldsymbol{\beta} \cdot \mathbf{S} = 0$) and has constant curvature, $K_R = \dfrac{|\dot{\boldsymbol{\beta}}|}{c} = \dfrac{\Omega_0}{c}$, [Eq. (20)]. The unique curve with zero torsion and constant curvature is a circle (the torsion and the curvature define unambiguously the shape of a curve apart from rotations and translations in space [12]). Thus, the trajectory of a free-particle in the co-moving frame is necessarily a circle perpendicular to the spin vector with radius $R_0 = \dfrac{1}{K_R} = \dfrac{c}{\Omega_0}$ (see Fig. 3ai). The same conclusion can be reached by direct integration of the equations of motion [Eq. (34a) and (34b)], as already illustrated in Sect. II.D [Eq. (18)]. The orbit radius $R_0$ is coincident with $R_C$, which for a time-crystal electron is the reduced Compton wavelength.

Interestingly, despite the continuous time translation symmetry of the system, in the co-moving frame there is a discrete time periodicity, so that the particle configuration is repeated over time, resulting in a time crystal state [4-6]. The time period of the circular trajectory is $T_0 = 2\pi/\Omega_0$. For a time-crystal electron with $\Omega_0 = m_e c^2/\hbar$ the duration of each time-crystal cycle is as short as $T_0 \sim 5 \times 10^{-20} s$. Curiously, as the orientation of the planar orbit is arbitrary, the family of circular time-crystal states forms a continuum, with each state determined by a different spin vector. This property is reminiscent of the fact that in quantum theory the ground state of a spin ½ particle is degenerate.

The particle trajectory in a generic inertial frame can be obtained by applying a Lorentz boost to the co-moving frame trajectory (a circle). It is however instructive to obtain an explicit expression for the trajectory by direct integration of the equations of motion for two particular cases.

In the first case (longitudinal motion), it is supposed that the center of mass velocity is aligned with the spin vector: $\mathcal{V} = \mathcal{V}\hat{\mathbf{z}}$ with $\mathbf{S} = S\hat{\mathbf{z}}$. In this situation, the constraint $\mathcal{V} \cdot \mathbf{S} = \mathbf{S} \cdot \boldsymbol{\beta}$ implies that $\mathcal{V} \cdot \boldsymbol{\beta} = \mathcal{V} \cdot \mathcal{V}$, so that Eq. (35) reduces to $\Omega_s = \Omega_0 \gamma_\mathcal{V}^{-2}$. In particular, for a longitudinal motion the frequency $\Omega_s$ (i.e., the normalized acceleration) is time-independent (see the green dashed line in Fig. 3b). Furthermore, due to Eq. (15) and $\mathcal{V} \cdot \boldsymbol{\beta} = \mathcal{V} \cdot \mathcal{V}$ the torsion of the $\mathbf{r}_0$ trajectory is also constant. Thus, the trajectory is necessarily a helix as this is only curve with a constant curvature ($K_R = \Omega_s/c$) and constant (nonzero) torsion.

This property can be confirmed by explicit integration of $\dot{\boldsymbol{\beta}} = \Omega_s \mathbf{S} \times \boldsymbol{\beta}$ which gives [see also Eq. (16)] $\boldsymbol{\beta}(t) = e^{(t-t_0)\omega_s \hat{\mathbf{z}} \times \mathbf{1}} \cdot \boldsymbol{\beta}(t_0)$ with $\omega_s = \Omega_s S$. For example, if $\boldsymbol{\beta}(t_0) = (\sin\theta_0 \cos\varphi_0, \sin\theta_0 \sin\varphi_0, \cos\theta_0)$ with $0 \le \theta_0 \le \pi$, the wave velocity is given by:

$$\boldsymbol{\beta}(t) = \left(\sin\theta_0 \cos(\omega_s(t-t_0)+\varphi_0),\ \sin\theta_0 \sin(\omega_s(t-t_0)+\varphi_0),\ \cos\theta_0\right). \tag{36}$$

with $|S| = 1/\sin\theta_0$. The corresponding $\mathbf{r}_0$ is easily found by explicit integration and evidently determines a helix with symmetry axis along $z$ (Fig. 3aii). The parameter $\theta_0$ has the same geometrical meaning as discussed previously, and is controlled by the center of mass velocity. In fact, from $\boldsymbol{\mathcal{V}} \cdot \boldsymbol{\beta} = \boldsymbol{\mathcal{V}} \cdot \boldsymbol{\mathcal{V}}$ one sees that $\mathcal{V} = \cos\theta_0$ and that the amplitude of the spin vector is $|S| = 1/\sin\theta_0 = \gamma_\mathcal{V}$.

In the second example, it is supposed that the center of mass velocity is perpendicular to the spin vector (transverse motion, with $\theta_0 = 90°$), so that $\boldsymbol{\mathcal{V}} = \mathcal{V}\hat{\mathbf{y}}$ with $\mathbf{S} = \hat{\mathbf{z}}$. Since the trajectory is planar it is possible to identify the wave velocity $\boldsymbol{\beta}$ with a complex number: $\boldsymbol{\beta}(t) = (\cos\varphi(t), \sin\varphi(t), 0) \sim e^{i\varphi(t)}$. Then, the equation $\dot{\boldsymbol{\beta}} = \Omega_s \mathbf{S} \times \boldsymbol{\beta}$ with $\mathbf{S} = \hat{\mathbf{z}}$ reduces to $\dot{\boldsymbol{\beta}} = i\Omega_s \boldsymbol{\beta}$ or equivalently to $\dot{\varphi} = \Omega_0 \dfrac{(1-\mathcal{V}\sin\varphi)^2}{1-\mathcal{V}^2}$ where Eq. (35) was used to evaluate $\Omega_s$. This equation can be easily integrated yielding:

$$-\mathcal{V} \frac{\cos(\varphi)}{1-\mathcal{V}\sin(\varphi)} + 2\gamma_\mathcal{V} \arctan\left(\gamma_\mathcal{V}\left(\tan\left(\frac{\varphi}{2}\right) - \mathcal{V}\right)\right) = \Omega_0 t. \tag{37}$$

The function $\varphi = \varphi(t)$ defined implicitly by the above equation is of the form $\varphi(t) = \varphi_P(t) + \dfrac{2\pi}{T}t$, where $\varphi_P(t)$ is a periodic function with period $T$ determined by the jump discontinuity of the arctan function, $2\gamma_\mathcal{V} \times \pi = \Omega_0 T$, i.e., $T = \gamma_\mathcal{V} 2\pi/\Omega_0$. In each time interval with duration $T$ the angle $\varphi$ increases by $2\pi$ and consequently $\boldsymbol{\beta}(t) \sim e^{i\varphi(t)}$ sweeps a full circle in the Bloch sphere. As illustrated in Fig. 3c, the angle $\varphi(t)$ (modulus $2\pi$) spends most of the oscillation period near $\varphi = \pi/2$, which corresponds to the *y*-axis. This confirms that the velocity vector sweeps the circle $\theta_0 = 90º$ in the Bloch sphere at a varying "speed", and justifies why there is a net translation of the particle towards the +*y*-direction (see Fig. 3aiii). Different from the first example, the acceleration $\Omega_s$ is not a constant of motion for a transverse trajectory (black line in Fig. 3b). In fact, the acceleration is minimized when $\mathcal{V}$ and $\boldsymbol{\beta}$ are parallel. In agreement with dynamical least action principle (Sect. II.E), the massless component is forced to move towards the direction $\mathcal{V}$, as near this direction $\boldsymbol{\beta}$ sweeps the great circle of the Bloch sphere more slowly (Fig. 3c).

The time average of $\boldsymbol{\beta}(t) \sim e^{i\varphi(t)}$ in one cycle can be computed explicitly using $\frac{1}{T}\int_0^T \boldsymbol{\beta}(t)dt = \frac{\Omega_0}{2\pi\gamma_\mathcal{V}} \int_{-\pi}^{\pi} e^{i\varphi} \frac{1}{\dot{\varphi}} d\varphi$. Substituting $\dot{\varphi} = \Omega_0 \frac{(1-\mathcal{V}\sin\varphi)^2}{1-\mathcal{V}^2}$ in the integral, one can show that $\frac{1}{T}\int_0^T \boldsymbol{\beta}(t)dt = i\mathcal{V} \to \mathcal{V}$. In other words, the time averaged $\boldsymbol{\beta}(t)$ is exactly coincident with the center of mass velocity:

$$\langle \boldsymbol{\beta} \rangle = \mathcal{V}. \tag{38}$$

The brackets $\langle ... \rangle$ stand for the operation of time averaging. Using the same ideas it is possible to evaluate the time average of the tensor $\boldsymbol{\beta} \otimes \boldsymbol{\beta}$. It is found that up to terms that are of order two in the center of mass velocity one has (with $\mathbf{S} = \hat{\mathbf{z}}$ the direction of the spin vector):

$$\langle \boldsymbol{\beta} \otimes \boldsymbol{\beta} \rangle \approx \frac{1}{2}(\mathbf{1} - \mathbf{S} \otimes \mathbf{S}) + o(\mathcal{V}^2). \tag{39}$$

## V. Generalized Lorentz force

The rest of the article is devoted to the study of a "time-crystal electron". I investigate different elementary interactions of the time-crystal electron with the electromagnetic field.

## A. The functions $L_{in}$ and $L_{out}$

As discussed in Sect. III.D, the time derivative of the massive momentum cannot be directly ruled by the classical Lorentz force due to the constraint (25). Such constraint implies that the force is given by the covariant expression (26), which is written in terms of the Lorentz scalars $L_{in}$ and $L_{out}$. Let us suppose that similar to the classical case the force depends linearly on the electric ($\mathbf{E}$) and magnetic ($\mathbf{B}$) fields. Furthermore, I consider that the force can have a nonclassical linear contribution arising from the massless momentum time derivative ($d\mathcal{P}/dt$). As already noted in Sect. III.B, the Lorentz scalar function $L_{in}$ should be even under the parity (P) and time-reversal (T) transformations. Thus, in the co-moving frame $L_{in}$ must be a linear combination of terms of the type $(\boldsymbol{\beta}\times\mathbf{S})\cdot\mathbf{V}$, $\boldsymbol{\beta}\cdot\mathbf{V}$, and $\mathbf{S}\cdot\mathbf{V}$, where $\mathbf{V}$ stands either for the electric field ($\mathbf{E}$) or magnetic field ($\mathbf{B}$) or the time derivative of the massless momentum ($d\mathcal{P}/dt$). I recall that $\boldsymbol{\beta}\times\mathbf{S}$, $\boldsymbol{\beta}$, and $\mathbf{S}$ form an orthonormal basis of space in the co-moving frame. The only 3 terms that are both P and T symmetric out of 9 possibilities are $(\boldsymbol{\beta}\times\mathbf{S})\cdot\mathbf{E}$, $\mathbf{S}\cdot\mathbf{B}$ and $(\boldsymbol{\beta}\times\mathbf{S})\cdot\dfrac{d\mathcal{P}}{dt}$. The coefficients of the linear combination must be constants to ensure that $L_{in}$ is also invariant under arbitrary rotations and translations of space. The previous discussion shows that a P and T symmetric linear coupling based on $\mathbf{E}$, $\mathbf{B}$ and $d\mathcal{P}/dt$ is necessarily of the form:

$$\frac{L_{in}}{R_C} = \left[(\boldsymbol{\beta}\times\mathbf{S})\cdot\frac{d\mathcal{P}}{dt}\right]_{co} + q_e g_L \left[(\boldsymbol{\beta}\times\mathbf{S})\cdot\mathbf{E} + \alpha_{RB}\mathbf{S}\cdot c\mathbf{B}\right]_{co}. \tag{40}$$

In the above, $g_L > 0$ and $\alpha_{RB}$ are coupling constants. The coefficient $\alpha_{RB}$ is a free-parameter, which will be taken equal to zero in all the examples of the article, but for completeness it is included in all the formulas. The fields are evaluated at the point $(\mathbf{r}_0, t)$. The coefficient $g_L$ will be related to the charge renormalization parameter $\theta_e$ [Eq. (24)] at a later point.

Proceeding in the same manner for $L_{out}$, taking into account that it should be odd under both the parity (P) and time-reversal (T) transformations, it can be shown that $L_{out}$ must be a linear combination of terms of the type: $(\boldsymbol{\beta}\times\mathbf{S})\cdot\mathbf{B}$, $\mathbf{S}\cdot\mathbf{E}$, and $\mathbf{S}\cdot\dfrac{d\mathcal{P}}{dt}$. It is simple to check that $\mathbf{S}\cdot\dfrac{d\mathcal{P}}{dt}$ vanishes in the co-moving frame. This leads to the result:

$$\frac{L_{\text{out}}}{R_{\text{C}}} = q_{\text{e}} g_{\text{L}} \left[ \alpha_{\text{SE}} \mathbf{S}\cdot\mathbf{E} + \alpha_{\text{SB}} (\boldsymbol{\beta}\times\mathbf{S})\cdot c\mathbf{B} \right]_{\text{co}}, \tag{41}$$

with $\alpha_{\text{SE}}, \alpha_{\text{SB}}$ some coupling coefficients to be determined. It will be proven in the following sections that to recover the classical Lorentz force and classical spin precession in the limit $\hbar \to 0$ it is necessary that $\alpha_{\text{SE}} = 1/2$, $\alpha_{\text{SB}} = 1$, and $g_{\text{L}} \approx 2$.

Taking into account the enunciated properties and that $L_{\text{in}}, L_{\text{out}}$ transform as Lorentz scalars, it can be shown that in a generic frame they are given by:

$$\frac{L_{\text{in}}}{R_{\text{C}}} = -m_{\text{e}}c\frac{\mathcal{E}\Omega_{\text{s}}}{E-\boldsymbol{\beta}\cdot c\boldsymbol{\pi}} + \gamma_{\mathcal{V}} g_{\text{L}} \left\{ \frac{1}{1-\boldsymbol{\beta}\cdot\mathcal{V}} \left[(\boldsymbol{\beta}-\mathcal{V})\times\tilde{\mathbf{S}}\right]\cdot\left[\mathcal{F}_{L,\text{e}} - \mathcal{V}(\mathcal{V}\cdot\mathcal{F}_{L,\text{e}})\right] + \alpha_{\text{RB}} \tilde{\mathbf{S}}\cdot\mathcal{F}_{L,\text{m}} \right\}, \tag{42a}$$

$$\frac{L_{\text{out}}}{R_{\text{C}}} = \gamma_{\mathcal{V}} g_{\text{L}} \left\{ \frac{1}{2}\tilde{\mathbf{S}}\cdot\mathcal{F}_{L,\text{e}} + \frac{1}{1-\boldsymbol{\beta}\cdot\mathcal{V}} \left[(\boldsymbol{\beta}-\mathcal{V})\times\tilde{\mathbf{S}}\right]\cdot\left[\mathcal{F}_{L,\text{m}} - \mathcal{V}(\mathcal{V}\cdot\mathcal{F}_{L,\text{m}})\right] \right\}, \tag{42b}$$

with $\mathcal{F}_{L,\text{e}} = q_{\text{e}}(\mathbf{E}+\mathcal{V}\times c\mathbf{B})$, $\mathcal{F}_{L,\text{m}} = q_{\text{e}}(c\mathbf{B}-\mathcal{V}\times\mathbf{E})$, $\tilde{\mathbf{S}} = \mathbf{S} - \mathcal{V}(\mathbf{S}\cdot\boldsymbol{\beta})$ and $\gamma_{\mathcal{V}}^2 = 1/(1-\mathcal{V}\cdot\mathcal{V})$. The generalized Lorentz force in a generic reference frame is given by Eq. (34d) with $L_{\text{in}}, L_{\text{out}}$ defined as above. In the co-moving frame, the force reduces to

$$\begin{aligned}\mathcal{F}_{\text{co}} = &-\Omega_{\text{s}}\frac{\mathcal{E}}{c}\boldsymbol{\beta}\times\mathbf{S} + q_{\text{e}} g_{\text{L}} \left[(\boldsymbol{\beta}\times\mathbf{S})\cdot\mathbf{E} + \alpha_{\text{RB}}\mathbf{S}\cdot c\mathbf{B}\right]\boldsymbol{\beta}\times\mathbf{S} \\ &+ q_{\text{e}} g_{\text{L}} \left[\frac{1}{2}\mathbf{S}\cdot\mathbf{E} + (\boldsymbol{\beta}\times\mathbf{S})\cdot c\mathbf{B}\right]\mathbf{S}\end{aligned}, \tag{43}$$

The terms in the first line give the in-plane force and the term in the second line gives the force along the spin vector direction. The electromagnetic fields in principle must include a self-contribution, e.g. the action of the charge positioned at the center of mass $\mathbf{r}_{\text{CM}}$ on the wave component $\mathbf{r}_0$. In this article, such self-contribution is ignored and $\mathbf{E},\mathbf{B}$ are treated as external fields. The term proportional to the wave energy $\mathcal{E}$ determines a self-force, as it is present even in the absence of external electromagnetic interactions (when $\mathbf{E}=0$ and $\mathbf{B}=0$). It will be seen in Sect. VI that the self-force is essential to bind the $\mathbf{r}_0$ and $\mathbf{r}_{CM}$ trajectories.

The dynamics of a time-crystal electron is fully determined by the system of equations (34) complemented with Eq. (42). The equations are Lorentz covariant. With the exception of the equation that controls the center of mass trajectory ($d\mathbf{r}_{CM}/dt = c\mathcal{V}_{CM}$), the remaining equations are time-reversal invariant. The equation $d\mathbf{r}_{CM}/dt = c\mathcal{V}(t_d)$ is not time-reversal invariant due to the time delay $t_d$. Interestingly, as $\mathbf{r}_{CM}$ is not directly coupled to the remaining equations, it follows that the dynamics of the wave-component of the electron (described by $\mathbf{r}_0, \mathbf{S}, \boldsymbol{\beta}, \mathcal{V}$) is exactly time-reversal invariant.

## *B. The classical limit*

Let us now focus on the dynamics of the center of mass. In the co-moving frame it is controlled by $m_e c \left( \dfrac{d\mathcal{V}}{dt} \right)_{co} = \mathcal{F}_{co}$ with the force given by Eq. (43). It is logical to impose that $\mathcal{F}$ reduces to the usual Lorentz force in the classical limit $\hbar \to 0$. In the $\hbar \to 0$ limit the spinning frequency $\Omega_s$ diverges to infinity and the period of the time-crystal cycle approaches zero. Thus, the classical limit can be found by averaging the force over a time-crystal cycle with the duration of the time period approaching zero $T_0 \to 0$ and with the radius of the orbit also approaching zero $R_0 \to 0$. This means that the time and space dependence of the fields can be ignored. In practice, the approximation $\hbar \to 0$ is acceptable when the (external) fields only vary on a spatial scale much larger than the reduced Compton wavelength and on a time scale much larger than the actual time-crystal cycle $T_0 \sim 2\pi\hbar/m_e c^2 \sim 5 \times 10^{-20} s$.

From the previous discussion, the coupling coefficients $g_L, \alpha_{SE}, \alpha_{SB}$ must be tuned to ensure that in the co-moving frame $\langle \mathcal{F}_{co} \rangle = q_e \mathbf{E}$. Note that if $\langle \mathcal{F}_{co} \rangle = q_e \mathbf{E}$ in the co-moving frame then $\langle \mathcal{F} \rangle = q_e (\mathbf{E} + \mathcal{V} \times c\mathbf{B})$ in a generic frame. It is implicit that the averaging is done with over a time-crystal cycle with $T_0 \to 0$. In the co-moving frame the time-averaged velocity vanishes ($\langle \boldsymbol{\beta} \rangle \approx 0$) and the spin vector varies slowly in time ($\langle \mathbf{S} \rangle \approx \mathbf{S}$) [see Eq. (38)]. From here, it follows that $\langle \boldsymbol{\beta} \times \mathbf{S} \rangle \approx 0$, $\langle (\boldsymbol{\beta} \times \mathbf{S}) \otimes \mathbf{S} \rangle \approx 0$, $\langle \mathbf{S} \otimes (\boldsymbol{\beta} \times \mathbf{S}) \rangle \approx 0$ and $\langle \mathbf{S} \otimes \mathbf{S} \rangle \approx \mathbf{S} \otimes \mathbf{S}$. In particular, one concludes that the terms that involve the self-force and the magnetic field in Eq. (43) do not contribute to the time-averaged (effective) force[†]. Noting that $\mathbf{1} = (\boldsymbol{\beta} \times \mathbf{S}) \otimes (\boldsymbol{\beta} \times \mathbf{S}) + \boldsymbol{\beta} \otimes \boldsymbol{\beta} + \mathbf{S} \otimes \mathbf{S}$, it is clear that $\langle (\boldsymbol{\beta} \times \mathbf{S}) \otimes (\boldsymbol{\beta} \times \mathbf{S}) \rangle = \mathbf{1} - \mathbf{S} \otimes \mathbf{S} - \langle \boldsymbol{\beta} \otimes \boldsymbol{\beta} \rangle$. Hence, using $\langle \boldsymbol{\beta} \otimes \boldsymbol{\beta} \rangle = \frac{1}{2}(\mathbf{1} - \mathbf{S} \otimes \mathbf{S})$ [Eq. (39)], it is readily found that: $\langle \mathcal{F}_{co} \rangle = g_L q_e \left( \frac{1}{2} \mathbf{E} + \left( \alpha_{SE} - \frac{1}{2} \right) \mathbf{S} \mathbf{S} \cdot \mathbf{E} \right)$. In order to recover the classical Lorentz force $\langle \mathcal{F}_{co} \rangle = q_e \mathbf{E}$ the coupling coefficients must be $g_L = 2$ and $\alpha_{SE} = 1/2$. In summary, the Lorentz scalars (40)-(41) with $g_L = 2$ and $\alpha_{SE} = 1/2$ ensure that the effective dynamics of a time-crystal electron is described by classical theory in the $\hbar \to 0$ limit.

## C. Spin precession

The effective dynamics of the spin vector can be determined with the same method as in the previous subsection. Substituting Eq. (41) into $\frac{d\mathbf{S}}{dt} = \frac{L_{out}}{\hbar} \boldsymbol{\beta}$ [Eq. (34c)], it is found that in the co-moving frame:

$$\left( \frac{d\mathbf{S}}{dt} \right)_{co} = \frac{q_e g_L}{m_e c} \boldsymbol{\beta} \left[ \alpha_{SE} \mathbf{S} \cdot \mathbf{E} + \alpha_{SB} (\boldsymbol{\beta} \times \mathbf{S}) \cdot c\mathbf{B} \right], \qquad (\mathcal{V} = 0 \text{ frame}). \qquad (44)$$

Neglecting the spatial gradients of the fields and taking into account that $\langle \boldsymbol{\beta} \otimes \boldsymbol{\beta} \rangle = \frac{1}{2}(\mathbf{1} - \mathbf{S} \otimes \mathbf{S})$ [Eq. (39)] and $\langle \boldsymbol{\beta} \otimes \mathbf{S} \rangle = 0$, one finds that in the classical limit $\hbar \to 0$ the spin vector dynamics is effectively controlled by:

$$\left( \frac{d\mathbf{S}}{dt} \right)_{co} \approx -\mathbf{S} \times \boldsymbol{\omega}_s. \qquad (45)$$

---

[†] Strictly speaking this is true only as a first order approximation. In Sect. VI, it shall be seen that the self-field originates a second order contribution to the force due to its fast variation in time.

with $\omega_s \approx \frac{-q_e}{m_e} \frac{g_L \alpha_{SB}}{2} \mathbf{B}$. Thus, consistent with the theory of Larmor precession, it is found that the spin vector of the time-crystal electron executes a precession motion about the (external) magnetic field. The spin precession frequency is $\omega_s$. In order that it can match the cyclotron frequency in the co-moving frame ($\omega_c = \frac{-q_e}{m_e} \mathbf{B}$) it is required that $\frac{g_L \alpha_{SB}}{2} = 1$ (a mismatch of the two frequencies due to the self-energy is discussed in the Sect. VI). Using $g_L = 2$ it follows that $\omega_c = \omega_s$ leads to $\alpha_{SB} = 1$, as anticipated in Sect. V.A.

### D. Stationary action principle for the center of mass and spin vector

Similar to the trajectory of the wave component, the dynamics of the center of mass and spin vector are also controlled by a stationary action principle, with $L_{in}, L_{out}$ playing a role analogous to the Lagrangian. For example, from Eq. (26) the in-plane center of mass acceleration in the co-moving frame is $\frac{d\mathcal{V}}{dt} = \frac{L_{in}}{\hbar} \boldsymbol{\beta} \times \mathbf{S}$. The vector $\hat{\mathbf{R}} = \boldsymbol{\beta} \times \mathbf{S}$ follows the same great circle in the Bloch sphere as the vector $\boldsymbol{\beta}$ apart from a 90º phase difference. The vector $\hat{\mathbf{R}}$ gives roughly the relative orientation of the wave component with respect to the center of mass: $\mathbf{r}_0 \approx \mathbf{r}_{CM} + R_C \hat{\mathbf{R}}$. Clearly, in each time crystal cycle the center of mass moves on average towards the direction $\hat{\mathbf{R}}$ that maximizes $L_{in}$.

Similarly, the motion of the spin vector is ruled by $\frac{d\mathbf{S}}{dt} = \frac{L_{out}}{\hbar} \boldsymbol{\beta}$ [Eq. (34c)]. From here and $\langle \boldsymbol{\beta} \rangle \approx 0$ [Eq. (38)] one sees that the spin vector in the co-moving frame will move on average towards the direction $\boldsymbol{\beta}$ that maximizes $L_{out}$, with $\boldsymbol{\beta}$ constrained to the great circle. Note the spin vector controls the out-of-plane motion of the time-crystal particle. As an illustration, consider the expression for $L_{out}$ given by Eq. (41). The direction $\boldsymbol{\beta}$ that maximizes $L_{out}$ is along the vector $q_e \alpha_{SB} (\mathbf{S} \times c\mathbf{B})$, supposing that the space gradients of the fields are negligible. This indicates that

-37-

the spin vector will move towards the direction $q_e \alpha_{SB}(\mathbf{S} \times c\mathbf{B})$, which corresponds to the spin precession motion discussed in Sect. V.C.

### E. Electric dipole approximation of the electromagnetic 4-potential

The spin frequency is a function of the electromagnetic 4-potential [Eq. (23)]. The electromagnetic 4-potential is gauge dependent. Thus, it is necessary to fix some gauge in order that the theory is unambiguously defined, e.g., the Lorenz gauge. In this article, for simplicity, I use a simple electric dipole approximation of the electromagnetic 4-potential.

Specifically, consider the formula for the spin frequency in the co-moving frame: $\Omega'_s = \left[ m_e c^2 + \theta_e q_e (\varphi' - c\boldsymbol{\beta}' \cdot \mathbf{A}') \right] / \hbar$. In the electric dipole approximation, the magnetic potential is ignored and the electric potential is simply replaced by $\varphi \to -\mathbf{R} \cdot \mathbf{E}$, with the position vector measured with respect to the center of mass coordinates taken as $\mathbf{R} \to R_C \boldsymbol{\beta} \times \mathbf{S}$. This approximation yields:

$$\Omega'_s \approx \left[ m_e c^2 - R_C \theta_e q_e (\boldsymbol{\beta}' \times \mathbf{S}') \cdot \mathbf{E}' \right] / \hbar. \tag{46}$$

## VI. Interactions with the electromagnetic field

### A. The self-force

Let us now discuss the role of the nonclassical force arising from the massless momentum time-derivative ($d\mathcal{P}/dt$). Ignoring the electromagnetic fields contribution ($\mathbf{E} = 0$, $\mathbf{B} = 0$), the force in the co-moving frame reduces to [Eq. (43)]: $\mathcal{F}_{co,self} = -\Omega_{s,co} \dfrac{\mathcal{E}_{co}}{c} \boldsymbol{\beta}' \times \mathbf{S}'$. This result can also be written in terms of the acceleration as:

$$\mathcal{F}_{self,co} = \mathcal{V}_{min} \frac{\hbar}{R_C} \dot{\boldsymbol{\beta}}, \quad \text{with} \quad \mathcal{V}_{min} = \frac{\mathcal{E}_{co}}{m_e c^2} \quad \text{(no external fields)}. \tag{47}$$

As seen, the self-force acting on the center of mass is proportional to the acceleration of the wave component. The parameter $\mathcal{V}_{min}$ is determined by the fraction of the wave energy ($\mathcal{E}_{co}$) relative to the rest mass energy ($m_e c^2$). As discussed in Sect. IV.C, $\mathcal{E}_{co}$ describes a self-interaction of the time-crystal with itself and is identified roughly with the self Coulomb potential: $\mathcal{E}_{co} \sim q_e \varphi_{co} = \alpha_e \hbar c / R_0$. Here $\varphi_{co}(\mathbf{r}_0)$ is the Coulomb potential created by the center-of-mass and $\alpha_e = q_e^2 / (4\pi\varepsilon_0 \hbar c) \approx 1/137$ the fine-structure constant of the electron. Using $\mathcal{E}_{co} \approx \hbar c / R_0$ one obtains the rough estimate $\mathcal{V}_{min} \sim \alpha_e$ (later it shall be seen that this formula overestimates $\mathcal{V}_{min}$), which indicates that $\mathcal{V}_{min} \ll 1$.

In the co-moving frame, the force can be written in terms of the acceleration as $\mathcal{F}_{co} = m_e c \left( \frac{d\mathcal{V}}{dt} \right)_{co}$. Hence, Eq. (47) implies that in the co-moving frame $\left( \frac{d\mathcal{V}}{dt} \right)_{co} = \mathcal{V}_{min} \dot{\boldsymbol{\beta}}_{co}$. Assuming $\mathcal{V}_{min} \ll 1$, it is acceptable to use $\frac{d\mathcal{V}}{dt} \approx \mathcal{V}_{min} \dot{\boldsymbol{\beta}}$ in an inertial frame where $|\mathcal{V}| \ll 1$. The integration of this equation yields $\mathcal{V} \approx \mathcal{V}_{min} \boldsymbol{\beta}$ when the mean velocity vanishes, $\langle \mathcal{V} \rangle \approx 0$. Thus, the normalized center of mass velocity amplitude is exactly $\mathcal{V}_{min}$, which justifies the used approximation. Hence, due to the self-force, the center-of-mass is never truly at rest and is dragged by the motion of the wave component. In the frame where $\langle \mathcal{V} \rangle \approx 0$, the root mean square (rms) velocity is on the order of $\mathcal{V}_{min} c$. Furthermore, integrating $\mathcal{V} \approx \mathcal{V}_{min} \boldsymbol{\beta}$ one finds that $\mathbf{r}_{CM} \approx \mathcal{V}_{min} (\mathbf{r}_0 - \mathbf{r}_c)$ with $\mathbf{r}_c$ is some integration constant. Thereby, in a quasi-rest frame ($\langle \mathcal{V} \rangle \approx 0$) the trajectory of the center of mass is approximately identical to the trajectory of the massless component, but scaled down by a factor of $\mathcal{V}_{min}$.

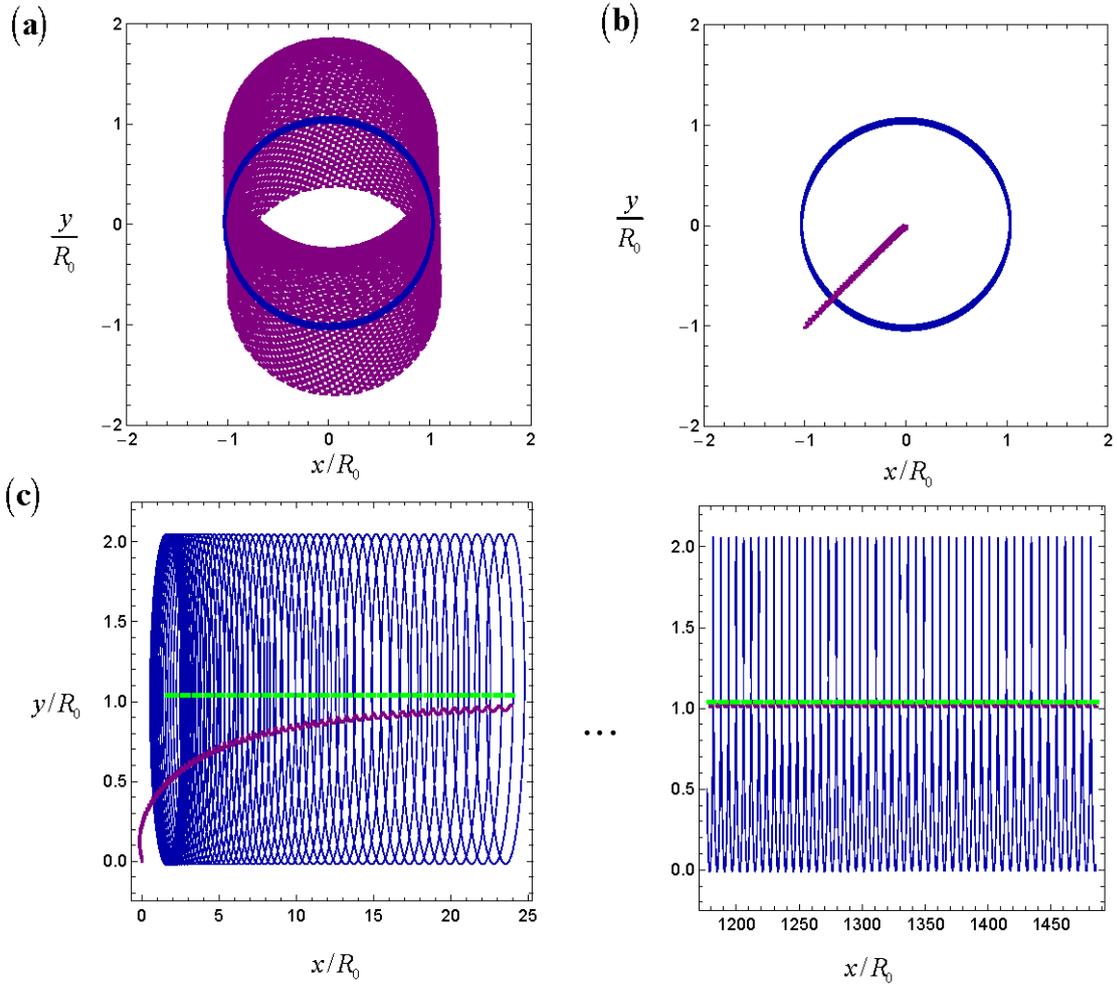

**Fig. 4 (a)-(b)** Trajectory of a time-crystal electron free from external interactions (100 cycles), subject only to the self-field described by $\mathcal{V}_{min} = 1/60$. The trajectories are planar with the spin vector directed along +z. Blue lines: trajectory of the wave-component $\mathbf{r}_0$. Purple lines: trajectory of the particle-component $\mathbf{r}_{CM}$. a) The trajectory of $\mathbf{r}_{CM}$ is enlarged by a factor of $1/\mathcal{V}_{min}$ so that it matches well the trajectory of the massless-component. b) Trajectory of the time-crystal electron when the particle-component position is initially offset from the geometrical center of the wave-component orbit. **(c)** Trajectory of a time-crystal electron subject to a static (extremely strong) electric field directed along the negative x-direction with constant intensity $q_e E_0 / m_e c = 3 \times 10^{-4} \Omega_0$. Left panel: illustration of the synchronization of the two trajectories in the initial time range $0 < t < 70 T_0$. Right panel: trajectory of the time-crystal electron in the time interval $490 T_0 < t < 570 T_0$. The static field accelerates the electron from a near zero velocity up to $\mathcal{V} = 0.7$ in the considered time range. The classical relativistic trajectory is represented by the green-dashed line and agrees very precisely with the actual trajectory.

Figures 4a and 4b show the trajectory of a time-crystal electron free from external interactions in the frame where $\langle \mathcal{V} \rangle = 0$. The equations of motion [Eq. (34) with the Lagrangian functions defined as in Eq. (42)] are numerically integrated using the 4$^{th}$ order Runge-Kutta method. The center of mass trajectory is determined by $d\mathbf{r}_{CM}/dt = c\mathcal{V}(t_d)$. As shown in Appendix B, this delay differential equation can be reduced to a standard differential equation and thereby integrated in a standard manner.

As seen in Fig. 4a, the two components of the electron follow circular-type orbits. For clarity, the center of mass trajectory was enlarged by a factor of $1/\mathcal{V}_{\min}$. Consistent with the previous discussion it nearly overlaps the trajectory of the wave component (the center of mass has a small but nonzero velocity along $y$ that prevents the perfect overlap of the two orbits during the 100 time crystal cycles of the simulation). The radius of the wave-component orbit is $R_0 = R_C$, whereas the radius of the particle-component orbit is $\mathcal{V}_{\min} R_C$. Note that for a $\mathcal{V}_{\min}$ on the order of the fine-structure constant $\alpha_e$ the radius $\mathcal{V}_{\min} R_C$ is on the order of classical electron radius ($\alpha_e R_C$).

The self-force effectively binds the center-of-mass and wave trajectories due to a drag effect. This is illustrated in Figure 4b, which corresponds to a scenario wherein the center of mass (purple curve) is initially offset from the geometrical center of the massless-component orbit (blue curve). As seen, as time passes the center-of-mass position slides to the center of the circular orbit. Without the self-force, the center of mass position $\mathbf{r}_{CM}$ would be independent of time in the rest frame. Clearly, a self-force with $\mathcal{V}_{\min} > 0$ is absolutely essential to bind the two particle components. The synchronization of the two trajectories also occurs in the presence of external fields. This is illustrated in Fig. 4c (left), which depicts the (planar) trajectory of a time crystal electron subject to a constant electric field directed along $-x$. The trajectory of the center of mass follows very precisely the classical relativistic trajectory of the electron (see the green line in Fig. 4c). The simulations are done with an extremely large electric field ($E_0 \approx 4 \times 10^{14}\ V/m$) to avoid long numerical simulations with many time crystal cycles.

## B. Spin gradient force

As shown in Sect.V.B, in the classical limit $\hbar \to 0$ the effective force satisfies $\langle \mathcal{F}_{co} \rangle = q_e \mathbf{E}$. In the following, non-classical corrections to the effective force due to field gradients are investigated. I focus on the effect of the magnetic field gradient (a similar analysis shows that the contribution from the electric field gradient vanishes).

Let us again consider the co-moving frame force given by Eq. (43). For $\hbar \to 0$ or for a homogeneous magnetic field the terms associated with the coefficients $\alpha_{SB}$ and $\alpha_{RB}$ do not contribute to the effective force. In presence of a magnetic field gradient, the term $\mathbf{B}(\mathbf{r}_0)$ has a fast varying in time component (even if the magnetic field itself varies "slowly" in time) due to the spinning motion of $\mathbf{r}_0$. In order to assess the impact of the field gradient, I use a Taylor expansion of the magnetic field with respect to the center of mass position: $\mathbf{B}(\mathbf{r}_0) \approx \mathbf{B}(\mathbf{r}_{CM}) + (\delta \mathbf{R} \cdot \nabla) \mathbf{B}$, with $\delta \mathbf{R} = \mathbf{r}_0 - \mathbf{r}_{CM}$. As the center of mass position changes slowly in time in the rest frame, the term $\mathbf{B}(\mathbf{r}_{CM})$ does not contribute to the effective force. Thus, the relevant additional piece of the force is (using $\alpha_{SB} = 1$):

$$\mathcal{F}_{spin,co} = q_e g_L c \left\{ \alpha_{RB} \mathbf{S} \cdot \mathbf{B} \left( \overleftarrow{\nabla} \cdot \delta \mathbf{R} \right) \boldsymbol{\beta} \times \mathbf{S} + \mathbf{B} \cdot \left( \left( \overleftarrow{\nabla} \cdot \delta \mathbf{R} \right) \boldsymbol{\beta} \times \mathbf{S} \right) \mathbf{S} \right\}. \tag{48}$$

Since in the co-moving frame the orbit of the time crystal electron is approximately circular, one can use the approximation $\delta \mathbf{R} \approx R_0 \boldsymbol{\beta} \times \mathbf{S}$ with $R_0 = R_C$ the radius of the circular orbit. Taking into account that $\langle (\boldsymbol{\beta} \times \mathbf{S}) \otimes (\boldsymbol{\beta} \times \mathbf{S}) \rangle = \frac{1}{2}(\mathbf{1} - \mathbf{S} \otimes \mathbf{S})$ and $\nabla \cdot \mathbf{B} = 0$, one readily finds that the time averaged gradient spin force is:

$$\langle \mathcal{F}_{spin,co} \rangle \approx q_e g_L c \frac{R_0}{2} \left\{ \alpha_{RB} \nabla (\mathbf{S} \cdot \mathbf{B}) - (\alpha_{RB} + 1) \mathbf{S} (\mathbf{S} \cdot \nabla)(\mathbf{S} \cdot \mathbf{B}) \right\}. \tag{49}$$

The magnetic field gradient is evaluated at the center of mass position. Note that in the classical limit ($\hbar \to 0$) the orbit radius approaches zero ($R_0 \to 0$) and thereby $\langle \mathcal{F}_{spin,co} \rangle$ vanishes.

However, for a finite $\hbar$, the spin-gradient force $\langle \mathcal{F}_{\text{spin,co}} \rangle$ is nontrivial and corresponds to a non-classical correction of the Lorentz force.

From a semi-classical standpoint, the gradient force is expected to be of the form $\nabla(\boldsymbol{\mu}_e \cdot \mathbf{B})$ with $\boldsymbol{\mu}_e$ the intrinsic magnetic moment of an electron. This is precisely the force that is characterized in the Stern-Gerlach experiment (with neutral atoms) [25]. It is possible to have a force with exactly this structure if $\alpha_{\text{RB}} = -1$. Independent of the value of $\alpha_{\text{RB}}$, the projection of the force along the spin vector can be written as:

$$\langle \mathcal{F}_{\text{spin,co}} \rangle \cdot \mathbf{S} \approx (\mathbf{S} \cdot \nabla)(\boldsymbol{\mu}_{\text{eq}} \cdot \mathbf{B}) . \tag{50}$$

where the equivalent magnetic dipole is $\boldsymbol{\mu}_{\text{eq}} = -q_e g_L c \frac{R_0}{2} \mathbf{S}$. The amplitude of the equivalent magnetic dipole moment ($|\boldsymbol{\mu}_{\text{eq}}| = 2\mu_B$) is on the order of the Bohr magneton ($\mu_B = |q_e|\hbar/2m_e$), comparable to quantum theory.

The equivalent dipole can be linked to the classical dipole moment ($\boldsymbol{\mu}_{\text{cl}} = \frac{1}{2} R_0 c q_e \mathbf{S}$) as $\boldsymbol{\mu}_{\text{eq}} = -g_L \boldsymbol{\mu}_{\text{cl}} \approx -2\boldsymbol{\mu}_{\text{cl}}$. It is remarkable that $\boldsymbol{\mu}_{\text{eq}}$ differs from $\boldsymbol{\mu}_{\text{cl}}$ by the correct Thomas factor of 2, but its orientation is seemingly the opposite of what it should be due to the leading minus sign. In particular, $\boldsymbol{\mu}_{\text{eq}}$ is parallel to the spin vector $\mathbf{S}$ rather than anti-parallel as it is classically expected. From the point of view of the time-crystal model this is not necessarily a problem. In fact, as outlined in Appendix A, the time-crystal electron is expected to behave as a point electric charge glued to a magnetic dipole. The magnetic dipole coupled to the Maxwell's equations $\boldsymbol{\mu}_e$ is a degree of freedom of the theory. If one takes $\boldsymbol{\mu}_e \sim \boldsymbol{\mu}_{\text{eq}}$, then Eq. (50) becomes fully consistent with the usual semi-classical approximation. On the other hand, consistency with quantum theory requires in addition that the spin up (spin down) quantum state is identified with a spin vector $\mathbf{S}$ anti-parallel (parallel) to the magnetic field. This particular aspect will be further developed in Sect. VI.D.

Finally, it is noted that in order to fully explain the Stern-Gerlach experiment it is necessary that there is a mechanism that acts to orient the spin vector in a direction that is either parallel or anti-parallel to the magnetic field (spatial quantization of the angular momentum). A discussion of such a mechanism is out of the intended scope of the article. When the spin vector is aligned with the magnetic field, the force $\langle \mathcal{F}_{spin,co} \rangle$ acts to deflect particles with different spin orientations in opposite directions, in agreement with the original experiment [25].

## C. Anomalous magnetic moment

In Sect. V.B it was shown that in the classical limit $\hbar \to 0$ the effective force in the co-moving frame is given by $\langle \mathcal{F}_{co} \rangle = q_e \mathbf{E}$ when $g_L = 2$ [see Eq. (34)]. In reality, the result $\langle \mathcal{F}_{co} \rangle = q_e \mathbf{E}$ is exact only when $\mathcal{V}_{min} = 0$, i.e., when the electromagnetic self-energy vanishes. A non-zero self-force induces fast variations in both $\mathcal{V}$ and $\mathbf{S}$ that originate second order contributions to the effective force $\langle \mathcal{F}_{co} \rangle$. Thus, in order to guarantee that $\langle \mathcal{F}_{co} \rangle = q_e \mathbf{E}$ when $\hbar \to 0$, it is necessary to slightly adjust the parameter $g_L$ to properly take into account the self-field corrections.

A rigorous perturbation analysis out of the scope of this article shows that *(i)* the effective force can be proportional to the electric field only if $g_L$ and the charge renormalization parameter $\theta_e$ [Eq. (24)] are linked as $g_L = \theta_e / 2$. *(ii)* in order that $\langle \mathcal{F}_{co} \rangle = q_e \mathbf{E}$, the charge renormalization coefficient must satisfy:

$$\frac{\theta_e}{4} \approx \frac{1}{1 - 2\mathcal{V}_{min}}. \quad (51)$$

This renormalization guarantees that $\langle \mathcal{F}_{co} \rangle = q_e \mathbf{E}$ up to corrections on the order of $\mathcal{V}_{min}^2$ and assumes $\alpha_{RB} = 0$.

It can be easily checked that the parameter $\theta_e$ can be set identical to unit if the particle is modeled by a bare charge $q_b$ given by $q_b = \sqrt{\theta_e} q_e \approx 2 q_e$, with $q_e$ the experimental charge. Thus, $\sqrt{\theta_e}$ relates the bare charge of a model with $\theta_e = 1$ with the experimental charge. Note that the electromagnetic fields are proportional to the charge, and hence the experimental fields are related to the bare fields ($\mathbf{E}_b$) as $\mathbf{E}_b = \sqrt{\theta_e}\mathbf{E}$, etc.

Moreover, it can be shown that the self-field induces a small shift in the spin precession frequency so that it becomes $\boldsymbol{\omega}_s \approx \frac{-q_e}{m_e}\frac{g_L}{2}\left(1-\frac{1}{2}\mathcal{V}_{min}\right)\mathbf{B}$. This formula is valid up to corrections on the order of $\mathcal{V}_{min}^2$. Thus, quite remarkably, the perturbation due to the self-electromagnetic energy leads to a mismatch between the spin precession and cyclotron frequencies, i.e., it leads to an electron $g_e$-factor determined by

$$\frac{g_e}{2} \equiv \frac{\omega_s}{\omega_c} \approx \frac{g_L}{2}\left(1-\frac{1}{2}\mathcal{V}_{min}\right) \approx 1 + \frac{3}{2}\mathcal{V}_{min}, \tag{52}$$

where Eq. (51) is used in the last identity. In particular, it follows that $\mathcal{V}_{min}$ controls the anomalous magnetic moment of the electron: $a_e = \frac{g_e}{2} - 1 \approx \frac{3}{2}\mathcal{V}_{min}$. The 1$^{st}$ order anomalous shift derived by Schwinger and Feynman with quantum electrodynamics is $a_e \approx \frac{\alpha_e}{2\pi}$. This suggests that in the present model $\mathcal{V}_{min} \approx \frac{\alpha_e}{3\pi} \sim 1/1300$. The ratio $\omega_s/\omega_c$ is independent of the spin vector orientation with respect to the magnetic field.

In order to illustrate the discussion, next I present a numerical simulation of the trajectory of a time-crystal electron under the influence of a static magnetic field oriented along the +z-direction: $\mathbf{B} = B_0\hat{\mathbf{z}}$ (Fig. 5). As previously noted, the characteristic time scale of the spinning motion is extremely short $T \sim 5\times 10^{-20}s$, about 10 orders of magnitude less than the cyclotron period determined by any realistic magnetic field. In order that the computational effort of the simulation is acceptable, the magnetic field is taken unrealistically large in the simulations so that the cyclotron frequency is $\omega_c = \Omega_0/10000$ (here, $\omega_c = -\dfrac{q_e}{\gamma m_e}B_0$ is the cyclotron frequency and $\Omega_0 = m_e c^2/\hbar$). The initial (center-of-mass) velocity is taken equal to $\mathcal{V} \approx 0.2$.

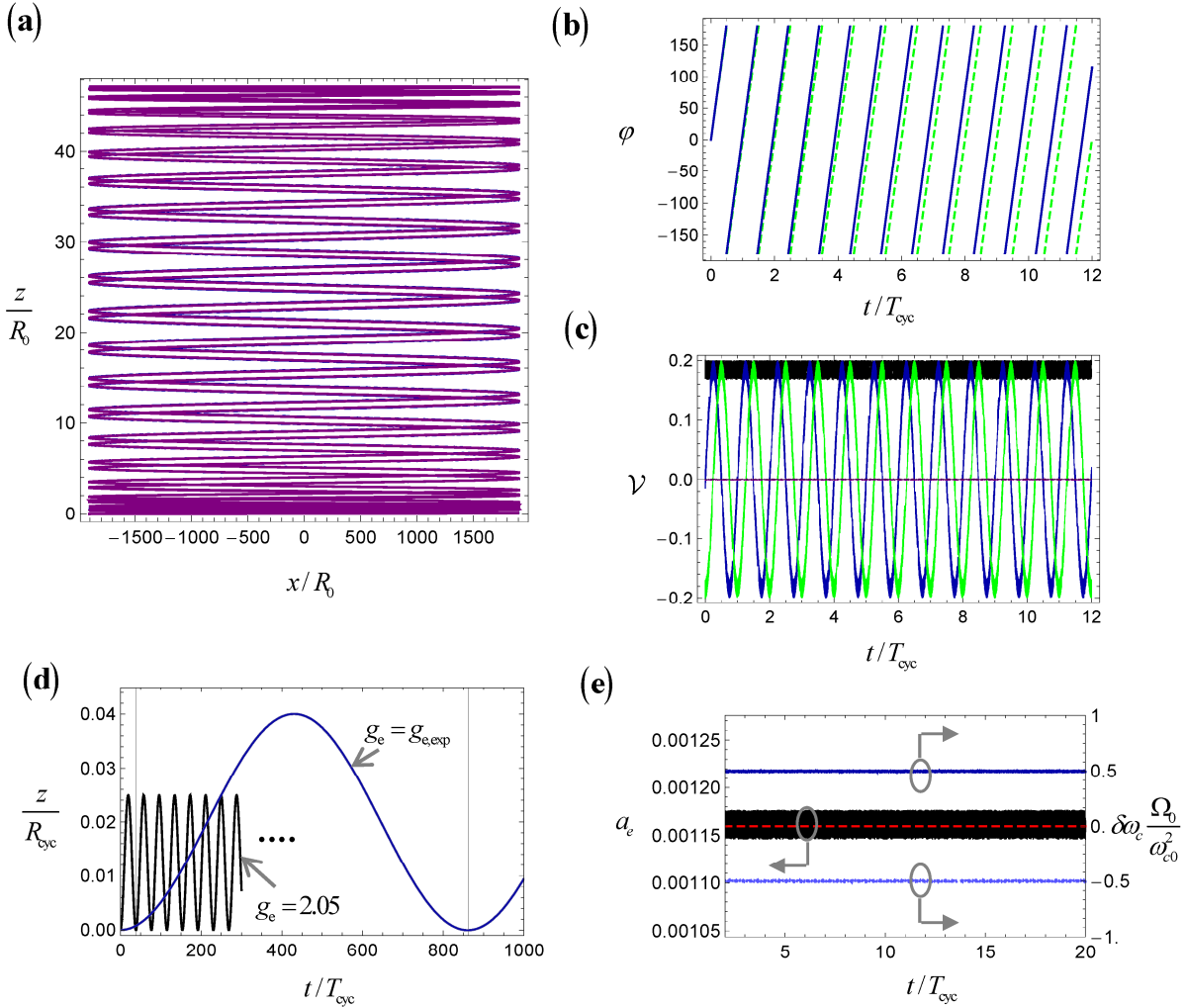

**Fig. 5** Trajectory of a time-crystal electron under the influence of a static magnetic field directed along +z, with the spin vector slightly tilted away from the +z axis by $\theta_S$. **a)** Trajectory in the *xoz* plane for $\mathcal{V}_{min} = 1/60$ and $\theta_S = 1°$. The blue and purple curves represent the trajectories of the wave and particle components. **b)** Angle swept by the in-plane spin vector (blue curve) and by the in-plane center-of-mass trajectory (green dashed curve) as a function of time normalized to the cyclotron period $T_{cyc} = 2\pi/\omega_c$. **c)** Normalized center of mass velocity as a function of $T_{cyc}$. The blue, green and purple curves represent $\mathcal{V}_x, \mathcal{V}_y, \mathcal{V}_z$, respectively, and the black curve represents the velocity amplitude. **d)** Axial displacement as a function of time for $\mathcal{V}_{min} = 1/60$ (black curve) and $\mathcal{V}_{min} \approx 1/1300$ corresponding to the experimental value of the anomalous magnetic moment (blue curve; this simulation was done using $\theta_S = 0.1°$, $\mathcal{V} = 0.02$, and $\omega_c = \Omega_0/60000$). **e)** Left axis: Anomalous magnetic moment as a function of time for $\mathcal{V}_{min} \approx 1/1300$.

The solid black curve shows the numerical result and the dashed red curve the experimental value. Right axis: Normalized shift of the cyclotron frequency $\delta\omega_c$ for $\mathcal{V}_{min} \approx 1/1300$. Solid (dark blue) curve: spin vector quasi-parallel to +z-direction; Dashed (light blue) curve: spin vector quasi-parallel to the –z-direction.

Figure 5b depicts the angle swept by the in-plane spin vector $\varphi_S = \arg(S_x + iS_y)$ (blue curve) and by the center of mass trajectory $\varphi_{CM} = \arg\{(x_{CM} - x_{CM0}) + i(y_{CM} - y_{CM0})\}$ (green dashed curve) as a function of time, for the case where the spin vector is tilted by 1° away from the +z-axis and the electromagnetic self-energy is such that $\mathcal{V}_{min} = 1/60$ (this corresponds to an electron g-factor of $g_e = 2.05$; the simulation is done with a relatively large $\mathcal{V}_{min}$ to illustrate more clearly the axial oscillation discussed below). Here, $(x_{CM}, y_{CM})$ are the in-plane coordinates of the center of mass and $(x_{CM0}, y_{CM0})$ are the coordinates of the circumference center. Clearly, the spin vector completes a full cycle faster than the center of mass, demonstrating that the spin precession frequency is larger than the cyclotron frequency. The anomalous magnetic moment can be numerically determined from the difference of slopes of $\varphi_S$ and $\varphi_{CM}$.

Figure 5c depicts the Cartesian components of the center of mass velocity as a function of time. The in-plane (*xoy*) components exhibit the expected sinusoidal variation in time. Interestingly, due to the self-force effect (Sect. VI.A), the amplitude of the total velocity (black curve) exhibits high-frequency oscillations around the mean value. The standard deviation of the velocity is on the order of $\mathcal{V}_{min}$.

Due to the extremely large cyclotron frequency the orbit of the time-crystal electron may deviate in a non-negligible way from the classical limit ($\hbar \to 0$ or equivalently $\Omega_0 \to \infty$) of a planar circular orbit. While the in-plane projection of the trajectory onto the *xoy* plane is nearly indistinguishable from a circumference (not shown), it turns out that when the initial spin vector is tilted by a non-trivial angle $\theta_S$ with respect to **B** the orbit is non-planar and exhibits an axial oscillatory motion along the *z*-direction. This is illustrated in Figs. 5a and 5d for the case $\mathcal{V}_{min} = 1/60$ (black curve, $\theta_S = 1°$) and in Fig. 5d for the case $\mathcal{V}_{min} \approx 1/1300$ (blue curve, $\theta_S = 0.1°$). As seen in Fig. 5d, the period of the in-out oscillatory axial motion (marked by the vertical gray lines) is roughly 40 cyclotron orbits for $\mathcal{V}_{min} = 1/60$ and roughly 862 cyclotron orbits for $\mathcal{V}_{min} \approx 1/1300$. The maximum axial displacement for $\mathcal{V}_{min} \approx 1/1300$ is about $0.04 R_{cyc}$ where $R_{cyc} \approx 1247 R_0$ is the radius of the cyclotron orbit. While the axial displacement is boosted by the giant magnetic field, which leads to a significant deviation of the trajectory with the respect to the classical case, the time period of the in-out oscillations has a rather fundamental origin, and its value is independent of $\theta_S$ and of the magnetic field amplitude for fields used in realistic experiments. This property is further discussed below. Figure 5e (left scale) plots the numerically calculated anomalous magnetic moment for $\mathcal{V}_{min} \approx 1/1300$, confirming that it agrees very precisely with the theoretical estimate of Eq. (52) (dashed red line in Fig. 5e). The high-frequency "noise" in the numerical calculation is mostly due to the unrealistically large value of $\omega_c / \Omega_0$.

Let us now discuss the physical origin of the in-out oscillatory axial motion (Figs. 5a and 5b). While in the limit $\hbar \to 0$ the time averaged force is $\langle \mathcal{F}_{co} \rangle = q_e \mathbf{E}$, for a finite $\hbar$ the averaged force may have additional terms. In the co-moving frame the averaged force must be written in terms of the vectors $\mathbf{E}, \mathbf{B}, \mathbf{S}$, as these are the only non-trivial vectors that vary slowly in time. The only way to construct a force from $\mathbf{E}, \mathbf{B}, \mathbf{S}$ that is even under a time reversal, odd under a parity transformation and that is linear in the fields is through a linear combination of the vectors $\mathbf{E}$ and $\mathbf{SS} \cdot \mathbf{E}$. Thus, for a finite $\hbar$ the averaged force in the co-moving frame may be assumed of the form $\langle \mathcal{F}_{co} \rangle = q_e (\mathbf{E} + \alpha_S \mathbf{SS} \cdot \mathbf{E})$ where $\alpha_S$ is some coefficient such that $\lim_{\hbar \to 0} \alpha_S = 0$. The force in an inertial frame wherein $\mathcal{V} \ll 1$ can be found with a Galilean transformation: $\langle \mathcal{F} \rangle \approx q_e \{(\mathbf{E} + \mathcal{V} \times c\mathbf{B}) + \alpha_S \mathbf{SS} \cdot (\mathbf{E} + \mathcal{V} \times c\mathbf{B})\}$. In particular, in the problem under analysis ($\mathbf{E} = 0$ and $\mathbf{B} = B_0 \hat{\mathbf{z}}$), the effective force reduces to $\langle \mathcal{F} \rangle \approx q_e \{\mathcal{V} \times c\mathbf{B} + \alpha_S \mathbf{SS} \cdot (\mathcal{V} \times c\mathbf{B})\}$. The first term is the familiar magnetic Lorentz force that originates the cyclotron motion. The second term is a correction due to the finite value of $\hbar$, or equivalently due to the finite value of $\omega_c/\Omega_0$. The correction vanishes in the limit $\omega_c/\Omega_0 \to 0$.

The amplitude of the correction term is modulated by the term $\mathbf{S}\cdot(\mathcal{V}\times c\mathbf{B}) = cB_0\mathcal{V}\sin(\varphi_S - \varphi_\mathcal{V})$ where $\varphi_S$ is the angle defined previously ($\varphi_S = \arg(S_x + iS_y)$) and $\varphi_\mathcal{V} = \arg(\mathcal{V}_x + i\mathcal{V}_y)$ is the angle swept by the in-plane center-of-mass velocity. It is clear that $\varphi_S - \varphi_\mathcal{V} = (\omega_s - \omega_c)t + \varphi_0$, and thereby the term $\mathbf{S}\cdot(\mathcal{V}\times c\mathbf{B})$ oscillates in time with frequency $\omega_s - \omega_c$. This is the origin of the axial oscillatory motion shown in Figs. 5a and 5d. The period of the axial oscillations is precisely:

$$T_{\text{in-out}} = \frac{2\pi}{\omega_s - \omega_c} = \frac{T_{\text{cyc}}}{a_e}, \tag{53}$$

with $T_{\text{cyc}} = 2\pi/\omega_c$ the period of the cyclotron orbit and $a_e = \frac{g_e}{2} - 1 = \frac{\omega_s}{\omega_c} - 1$ the anomalous magnetic moment. Remarkably, the period of the axial oscillations is independent of the spin vector orientation and is independent of the value of $\omega_c/\Omega_0$ (especially for sufficiently small $\omega_c/\Omega_0$). Thus, the axial oscillations occur even for small (and realistic) field amplitudes. For weaker fields ($\omega_c/\Omega_0 \to 0$) the amplitude of the axial displacement becomes negligible (as compared to $R_{\text{cyc}}$) because $\alpha_S \to 0$ in this limit. Thus, the time-crystal model of the electron provides an intuitive picture for the possible physical origin of the anomalous magnetic moment of the electron, relating it to the axial oscillations of the cyclotron motion. For $a_e \approx \alpha_e/2\pi$ the period of the axial oscillations is $T_{\text{in-out}} \approx 862 T_{\text{cyc}}$, i.e. it takes about 862 cyclotron orbits to complete a full axial oscillation (blue curve in Fig. 5d). Remarkably, the proposed mechanism is fully consistent with the way that the anomalous magnetic moment is experimentally determined using a Penning trap: the anomalous shift is measured by detecting a resonance for an axial excitation with frequency $\omega_s - \omega_c$ [26].

## D. Spin-dependent center-of-mass energy shift

A different nonclassical correction of the orbit due to the finite value of $\hbar$ is related to the cyclotron frequency $\omega_c$, which suffers a tiny spin dependent shift with respect to the classical value $\omega_{c0} \approx -\frac{q_e}{m_e} B_0$. It turns out that when the spin vector is parallel to the magnetic field the actual cyclotron frequency $\omega_c$ is slightly larger than the expected theoretical value $\omega_{c0}$. Similarly, when the spin vector is anti-parallel to the magnetic field the actual cyclotron frequency is slightly smaller than $\omega_{c0}$. The frequency shift $\delta\omega_c = \omega_c - \omega_{c0}$ may be attributed to some kind of rotational drag induced by the fast spinning motion of the wave-component.

From a different point of view, the shift $\delta\omega_c$ may be attributed to a shift of the (center-of-mass) particle rest energy, $c^2 m_{ef} = c^2 m_e + \delta E$, with the energy shift $\delta E$ dependent on the orientation of the spin vector. Clearly, one has $\omega_c \approx \omega_{c0}\left(1 - \frac{\delta E}{m_e c^2}\right)$, so that $\delta\omega_c / \omega_{c0} \approx -\delta E / m_e c^2$.

A perturbation analysis shows that –when the spin vector is either parallel or anti-parallel to the magnetic field– the shift $\delta\omega_c$ is given by $\delta\omega_c / \omega_{c0} \approx \hbar\boldsymbol{\omega}_c \cdot \mathbf{S} / (2 m_e c^2) \approx \pm 0.5 \omega_{c0} / \Omega_0$. This result is supported by the numerical simulations, see Fig. 5e, right scale calculated for $\mathcal{V}_{min} \approx 1/1300$. The frequency shift is to leading order independent of the value of $\mathcal{V}_{min}$. For realistic field amplitudes the shift $\delta\omega_c$ is negligibly small. The corresponding rest mass shift is $\delta E = -\hbar\boldsymbol{\omega}_c \cdot \mathbf{S} / 2$.

Interestingly, quantum mechanics predicts an identical spin-dependent energy shift. Specifically, for the spin up state the energy of an electron is shifted by $\hbar\omega_c / 2$, whereas for the spin down state it is shifted by $-\hbar\omega_c / 2$. In order that the energy shift $\delta E = -\hbar\boldsymbol{\omega}_c \cdot \mathbf{S} / 2$ can match the quantum mechanics result it is necessary that a spin vector parallel (anti-parallel) to the magnetic field corresponds to the usual spin down (spin up) quantum state. Thus, as already anticipated in the discussion of the spin-gradient force, the mapping between the spin vector of the time-crystal electron and the spin state in quantum theory must be the opposite of what could be expected. Due to the same reason, the link between the angular momentum and the intrinsic magnetic moment is also the opposite of quantum mechanics[‡].

## VII. Summary

In this article, I revisited de Broglie's old idea of an internal clock in the context of the more modern concept of time crystals. Remarkably, it was shown that the singular case of classical mechanics corresponding to a particle with zero mass yields a dynamical model that describes time-crystal states. It was found that time-crystal particles are characterized by an incessant spinning motion, and thereby have an intrinsic spin angular momentum. Furthermore, it was shown that a time crystal particle is characterized by a spin 4-vector that has a purely kinematic origin.

A time-crystal particle is characterized by two trajectories: the center of mass trajectory that moves with a speed less than $c$, and the wave trajectory that probes the nearby space at the speed of light. In particular, a time-crystal electron is formed by two inseparable components: (i) the particle-component that transports the charge and (ii) the wave-component that whirls around the "particle" and generates the spin and an intrinsic angular momentum. The trajectory of the particle is fully controlled by the trajectory of the wave, reminiscent of the pilot-wave theory of de Broglie and Bohm. Furthermore, in the proposed model the particle mass is an emergent property, in the sense that it originates from the fact that the center of mass frame speed is necessarily less than $c$.

A time-crystal particle is characterized by three fundamental 4-vectors. The massless energy-momentum 4-vector $(\mathcal{E}, \mathcal{P})$ determines the energy and momentum of the "wave". The spin 4-vector $(\mathbf{S} \cdot \boldsymbol{\beta}, \mathbf{S})$ determines orientation in space of the spinning motion. Finally, the massive energy-momentum 4-vector $(E,\boldsymbol{\pi})$ is the analogue of the classical energy-momentum. There is no strict conservation law associated with $(E,\boldsymbol{\pi})$, but the effective dynamics of the center of mass is consistent with classical theory when all the relevant frequencies are much less than the frequency of the spinning motion.

---

[‡] In principle, in deterministic models the angular momentum does not need to be linked to the intrinsic magnetic momentum in the same way as in quantum theory. In fact, quantum theory does not make independent predictions for the two quantities as they are measured exactly in the same way. A deterministic theory that reproduces the predictions of quantum mechanics needs evidently to predict the same magnetic moment properties (as the magnetic moment can originate macroscopic magnetism in solids [27], etc), the same energy shifts, the same deflections in space, etc, but not necessarily the same angular momentum, as the latter is not associated with any effect that does not relate to a manifestation of the intrinsic magnetic moment.

The trajectory of a time-crystal particle is controlled by a dynamical least action principle. The massless-component dynamically probes the nearby space and the particle moves on average towards the direction of space that minimizes the action integral. The electromagnetic self-energy originates a self-force that keeps the wave and particle components tightly attached. A time-crystal electron has a size on the order of the reduced Compton wavelength. The self-field interaction does not lead to singularities or infinite fields, different from both classical and quantum theories.

The model predicts the precession of the spin vector about a static magnetic field. Moreover, it relates the mismatch between the spin precession frequency and the cyclotron frequency –which is at the origin of the anomalous magnetic moment $a_\mathrm{e}$– with a manifestation of the electromagnetic self-energy. The time-crystal model predicts that the difference between $\omega_\mathrm{c}$ and $\omega_\mathrm{s}$ results in an axial oscillatory motion, and this picture appears to be supported by experiments [26]; the number of cyclotron orbits required to complete a full axial oscillation is $1/a_\mathrm{e} \approx 862$. The time-crystal model also predicts other known non-classical effects such as a spin-dependent **B**-field gradient force and a spin-dependent rest-energy shift.

The developed ideas suggest that time-crystal models of particles may capture some of the peculiar features of the quantum world, and thereby that it may be worthwhile to explore further these models.

**Acknowledgements:** This work is supported in part by the IET under the A F Harvey Engineering Research Prize, by the Simons Foundation under the award 733700 (Simons Collaboration in Mathematics and Physics, "Harnessing

# Appendix A: Maxwell field coupling

In the following, I discuss how a "time-crystal electron" may be coupled to the Maxwell's equations. Let us start with the electric charge density $\rho$. As there are two trajectories associated with the particle, $\mathbf{r}_0(t)$ and $\mathbf{r}_{CM}(t)$, there are two possibilities for a point-charge coupling, (i) $\rho = q_e \delta(\mathbf{r} - \mathbf{r}_0)$ or (ii) $\rho = q_e \delta(\mathbf{r} - \mathbf{r}_{CM})$, with $q_e = -e$ the electron charge. Due to the charge continuity equation, the electric current density associated with the option (i) is $\mathbf{j} = c\boldsymbol{\beta} q_e \delta(\mathbf{r} - \mathbf{r}_0)$. Such coupling leads to a very fundamental problem: as the acceleration $\dot{\boldsymbol{\beta}}$ cannot vanish (postulate P2), a particle described by (i) would continuously emit light. Furthermore, even more problematic, it can be checked that when $|\boldsymbol{\beta}| = 1$ the fields radiated by a current $\mathbf{j} = c\boldsymbol{\beta} q_e \delta(\mathbf{r} - \mathbf{r}_0)$ may have infinitely large amplitude [14]. Thus, the coupling (i) is not feasible, and one is left with the option (ii). For the option (ii), the electric current density is of the form $\mathbf{j} = c\boldsymbol{\mathcal{V}}_{CM} q_e \delta(\mathbf{r} - \mathbf{r}_{CM})$, analogous to a standard massive point charge.

Due to the intrinsic angular momentum of the electron, it is logical to consider as well a magnetic-dipole coupling. Specifically, it is suggested that in the frame co-moving with $\mathbf{r}_{CM}$ (i.e., in the inertial frame where $\boldsymbol{\mathcal{V}}_{CM} = 0$) the time-crystal electron is equivalent to a point electric charge $q_e$ "glued" to a magnetic dipole $\boldsymbol{\mu}_{e,co}$. In analogy with quantum theory, the magnetic dipole is assumed proportional to the (time-delayed) spin vector:

$$\boldsymbol{\mu}_{e,co} \equiv \boldsymbol{\mu}_e \big|_{\mathcal{V}_{CM}=0} = \mu_e \mathbf{S}_{CM} \big|_{\mathcal{V}_{CM}=0}, \tag{A1}$$

where $\mathbf{S}_{CM}=\mathbf{S}(t_d)$. From quantum theory and from the theory of magnetism in solids [27], the expected magnitude for $\mu_e$ is $|\mu_e| \sim \frac{g_e}{2}\sqrt{3}\mu_B \sim 2\mu_B$ with $\mu_B = \frac{|q_e|\hbar}{m_e 2}$ the Bohr magneton and $g_e$ the electron g-factor.

As discussed in the supplementary information [13, Sect. F], $(\boldsymbol{\mathcal{V}}_{CM}\cdot\mathbf{S}_{CM}, \mathbf{S}_{CM})$ transforms as a 4-vector and is a time-delayed version of the 4-vector $(\mathbf{S}\cdot\boldsymbol{\beta},\mathbf{S})$. Due to the constraints (27), the spin vector satisfies $\mathbf{S}_{CM}\cdot\mathbf{S}_{CM} - (\boldsymbol{\mathcal{V}}_{CM}\cdot\mathbf{S}_{CM})^2 = 1$. Thus, in the frame where $\boldsymbol{\mathcal{V}}_{CM}=0$ the spin vector has unit norm: $|\mathbf{S}_{CM}|=1$. This property shows that $|\boldsymbol{\mu}_e|_{\mathcal{V}_{CM}=0} = |\mu_e|$, so that the magnetic dipole moment is a constant in the frame co-moving with $\mathbf{r}_{CM}$.

The dipole $\boldsymbol{\mu}_{e,co}$ is seen in a generic frame as a magnetic dipole ($\boldsymbol{\mu}_e$) glued to an electric dipole ($\mathbf{p}_e$). The vector amplitudes of the two dipoles are given by:

$$\boldsymbol{\mu}_e = \left(\mathbf{1} - \frac{\gamma_{CM}}{\gamma_{CM}+1}\boldsymbol{\mathcal{V}}_{CM}\otimes\boldsymbol{\mathcal{V}}_{CM}\right)\cdot\boldsymbol{\mu}_{e,co}, \qquad \mathbf{p}_e = \frac{1}{c}\boldsymbol{\mathcal{V}}_{CM}\times\boldsymbol{\mu}_{e,co}, \qquad (A2)$$

with $\gamma_{CM} = 1/\sqrt{1-\boldsymbol{\mathcal{V}}_{CM}\cdot\boldsymbol{\mathcal{V}}_{CM}}$. The above formula is a consequence of the relativistic transformation of the electric polarization and magnetization vectors (which transform in the same manner as the magnetic and electric fields, $\mathbf{B}$ and $\mathbf{E}$, respectively) [14] and of $\mathbf{p}_e|_{\mathcal{V}_{CM}=0} = 0$. The electric dipole term may be regarded as some sort of electric "Röntgen current" induced by the motion of the magnetic dipole (see Ref. [28]).

Using $\boldsymbol{\mu}_{e,co} = \mu_e \mathbf{S}_{CM}|_{\mathcal{V}_{CM}=0}$ in Eq. (A2) and taking into account that $(\boldsymbol{\mathcal{V}}_{CM}\cdot\mathbf{S}_{CM}, \mathbf{S}_{CM})$ is a 4-vector, one can readily show that the dipole moments can be written as:

$$\boldsymbol{\mu}_e = \mu_e(\mathbf{1} - \boldsymbol{\mathcal{V}}_{CM}\otimes\boldsymbol{\mathcal{V}}_{CM})\cdot\mathbf{S}_{CM}, \qquad \mathbf{p}_e = \mu_e\frac{1}{c}\boldsymbol{\mathcal{V}}_{CM}\times\mathbf{S}_{CM}. \qquad (A3)$$

From the previous analysis, the electric current density and electric charge density coupled to the Maxwell's equations in a generic inertial frame must be of the form:

$$\mathbf{j} = \underbrace{q_e c \boldsymbol{\mathcal{V}}_{CM} \delta(\mathbf{r} - \mathbf{r}_{CM})}_{\text{point charge}} + \underbrace{\frac{d}{dt}\left[\mathbf{p}_e \delta(\mathbf{r} - \mathbf{r}_{CM})\right]}_{\text{electric dipole}} + \underbrace{\nabla \times \{\boldsymbol{\mu}_e \delta(\mathbf{r} - \mathbf{r}_{CM})\}}_{\text{magnetic dipole}}. \tag{A4a}$$

$$\rho = \underbrace{q_e \delta(\mathbf{r} - \mathbf{r}_{CM})}_{\text{point charge}} - \underbrace{(\mathbf{p}_e \cdot \nabla)\delta(\mathbf{r} - \mathbf{r}_{CM})}_{\text{electric dipole}}. \tag{A4b}$$

The electric current and charge densities satisfy the continuity equation $\nabla \cdot \mathbf{j} + \partial_t \rho = 0$. The proposed current density and charge density transform as a 4-vector as it should be. It is relevant to note that even though $\mathbf{p}_e \big|_{\mathcal{V}_{CM}=0} = 0$, the contribution of the electric dipole to the current density may be nontrivial in the frame co-moving with $\mathbf{r}_{CM}$ because $d\mathbf{p}_e / dt$ does not need to vanish.

## Appendix B: Integration of the center of mass trajectory

Equation (32) is a delay differential equation. It can be reduced to a ordinary differential equation by introducing a time advanced instant $t_a$ defined by $t_a = t + \frac{1}{c}|\mathbf{r}_{CM}(t_a) - \mathbf{r}_0(t)|$. Then, from the definition of the time delayed instant introduced in Sect. IV.A one has $t_a = (t_a)_d + \frac{1}{c}|\mathbf{r}_{CM}(t_a) - \mathbf{r}_0((t_a)_d)|$. This proves that $(t_a)_d = t$. Hence, the dynamics of $\mathbf{r}_{CM}(t_a)$ is determined by:

$$\frac{d}{dt}\left[\mathbf{r}_{CM}(t_a)\right] = c\boldsymbol{\mathcal{V}}((t_a)_d)\frac{dt_a}{dt} = c\boldsymbol{\mathcal{V}}(t)\frac{dt_a}{dt}. \tag{B1}$$

Differentiating both sides of $t = t_a - \frac{1}{c}|\mathbf{r}_{CM}(t_a) - \mathbf{r}_0(t)|$ with respect to time one can readily show that $\frac{dt_a}{dt} = \frac{1 + \hat{\mathbf{R}} \cdot \boldsymbol{\beta}(t)}{1 + \hat{\mathbf{R}} \cdot \boldsymbol{\mathcal{V}}(t)}$ with $\hat{\mathbf{R}} = \frac{\mathbf{r}_0(t) - \mathbf{r}_{CM}(t_a)}{|\mathbf{r}_0(t) - \mathbf{r}_{CM}(t_a)|}$. This proves that the time advanced center of mass trajectory can be found by solving the ordinary differential equation:

$$\frac{d}{dt}\left[\mathbf{r}_{CM}(t_a)\right] = c\boldsymbol{\mathcal{V}}(t)\frac{1 + \hat{\mathbf{R}} \cdot \boldsymbol{\beta}(t)}{1 + \hat{\mathbf{R}} \cdot \boldsymbol{\mathcal{V}}(t)}. \tag{B2}$$

The numerical simulations determine directly the time advanced trajectory, which is the quantity plotted in the figures.